\documentclass[epj,final,numbook]{svjour}
\usepackage{epsfig}
\begin{document}
\title{Luttinger liquids with boundaries: Power-laws and energy scales}
\author{V.\ Meden\inst{1,2}, W.\ Metzner\inst{1}, 
U.\ Schollw\"ock\inst{3}, O.\ Schneider\inst{1}, T.\ Stauber\inst{2}, 
and K.\ Sch\"onhammer\inst{2}}
\institute{Institut f\"ur Theoretische Physik C, 
RWTH Aachen, D-52056 Aachen, Germany \and 
Institut f\"ur Theoretische Physik, Universit\"at
G\"ottingen, Bunsenstr.\ 9, D-37073 G\"ottingen, Germany  \and
Sektion Physik, Universit\"at M\"unchen, Theresienstr.\ 37,
D-80333 M\"unchen, Germany}
\date{February 15, 2000}
\abstract{We present a study of the one-particle spectral properties for 
a variety of models of Luttinger liquids with open boundaries. 
We first consider the Tomonaga-Luttinger model using bosonization.
For weak interactions the boundary exponent of the power-law 
suppression of the spectral weight close to the chemical potential is 
dominated by a term linear in the interaction. 
This motivates us to study the spectral properties also within 
the Hartree-Fock approximation. It already gives power-law behavior 
and qualitative agreement with the exact spectral function. 
For the lattice model of spinless fermions and the
Hubbard model we present numerically exact results obtained using the
density-matrix renormalization-group algorithm. 
We show that many aspects of the behavior of the spectral function 
close to the 
boundary can again be understood within the Hartree-Fock 
approximation. 
For the repulsive Hubbard model with interaction $U$ the spectral 
weight is enhanced in a large energy range around the chemical 
potential. At smaller energies a power-law suppression, as predicted 
by bosonization, sets in. 
We present an analytical discussion of the crossover and show that
for small $U$ it occurs at energies exponentially (in $-1/U$) 
close to the chemical potential, i.e.\ that bosonization only 
holds on exponentially small energy scales. We show that such a crossover 
can also be found in other models.
\PACS{
      {71.10.-w}{Theories and models of many electron systems}   \and
      {71.10.Pm}{Fermions in reduced dimensions}   
     }}
\titlerunning{Luttinger liquids with boundaries: Power-laws and energy
  scales}
\authorrunning{V.\ Meden et al.}
\maketitle
\section{Introduction}
Theoretically it is well established that interacting fermi\-ons in one
spatial dimension do not obey Fermi liquid theory\cite{Voit}. 
The generic low-energy 
physics of  one-dimensional (1D) metallic fermions with repulsive
interaction can be described by Luttinger liquid (LL)
theory\cite{Voit,Tomonaga,Luttinger,Mattis,Solyom,Haldane}. For various
correlation functions it predicts asymptotic power-law behavior with 
exponents which, for spin rotational invariant models, can be expressed 
in terms
of a single parameter $K_{\rho}$. The Luttinger liquid parameter
$K_{\rho}$ depends on details of the model considered, e.g.\ the
interaction, filling factor, and one-particle dispersion\cite{Haldane,Voit}.  
It has been determined for many different models
of 1D correlated electrons using either analytical or numerical
techniques\cite{Schulz,Qin,Voit}. 
While the basic understanding of LL behavior emerged in
the study of 1D systems with periodic boundary conditions (PBC), 
the theoretical expectation that LL's with PBC including 
impurities scale to chains with open ends\cite{MattisII,KF} led to several
studies of models with hard walls,
usually called ``open'' (or ``fixed'') boundary conditions
(OBC)\cite{FG,EMJ,WVP,VWG,BBHN}. 
One way to experimentally verify  the predicted LL behavior
is to probe the one-particle properties using high resolution
photoemission spectroscopy. For the spectral
function $\rho(\omega)$ entering the description of angular integrated
photoemission and energies asymptotically close to the chemical
potential $\mu$,
LL theory predicts power-law suppression of the bulk spectral weight 
with an exponent 
\begin{eqnarray}
\alpha = (K_{\rho} + K^{-1}_{\rho} -2)/(2 z),
\label{alphabulk}
\end{eqnarray}
where $z=1$ for spinless fermions and $z=2$ for spin $1/2$-fermions. 

For the translational invariant system the scattering processes which
dominate the low-energy physics can be classified as forward,
backward, and um\-klapp scattering\cite{Solyom}. If the model parameters
are such that backward or umklapp scattering become
relevant in the 
renormalization-group (RG) sense both processes can drive the system
into a gapped non-LL phase. 
In lattice models and for commensurate filling factors 
umklapp scattering becomes relevant at a critical value of  
$K_{\rho}$ which dependents on the filling. For the lattice models
considered here we will always chose the interaction and filling such
that umklapp scattering remains irrelevant.
Standard RG arguments can be used\cite{Solyom}
to show that for repulsive interactions the $2k_F$-scattering part 
of the two-particle interaction
(usually called ``$g_1$-interaction'')  
scales to zero, where $k_F$ denotes the Fermi wave vector.
Then the Tomonaga-Luttinger (TL) 
model\cite{Tomonaga,Luttinger} describes the
generic low-energy LL physics. It only contains the scattering processes
with small momentum transfer, which can be written as a quadratic form
in bosonic density operators of left and right moving fermions. This
bosonization of the Hamiltonian is one of the ways to exactly solve
the TL model with PBC\cite{Voit,Bosintr}. 
For OBC we do not have momentum conservation and a general two-body 
interaction leads to a variety of different scattering vertices 
(see Sec.\ \ref{OBC}). They depend on different combinations
(differences and sums) of the four external quantum numbers and cannot
simply be parameterized by the ``momentum transfer'' as for PBC. 
Only a few of the vertices can be written as  quadratic forms in boson 
operators similar to the case of PBC\cite{Tomonaga,Solyom}. In some of 
the previous publications on the open boundary
problem\cite{FG,EMJ} it was tacitly assumed that RG arguments similar
to the bulk case can be applied to show that the remaining 
vertices of the system without translational invariance scale to zero. 
Therefore a quadratic form in boson operators was used to
describe the electron-electron interaction.  Then
it is straightforward to calculate correlation functions\cite{FG,EMJ}. 
In Sec.\ \ref{OBC} we present a detailed discussion of 
bosonization for the case of OBC taking all the scattering
processes into account. We explicitly demonstrate that the above 
assumption is justified for an
interaction which is long range in real space, considered by Tomonaga
for PBC\cite{Tomonaga}.     
In Refs.\ \cite{FG} and \cite{EMJ} it was shown that the 
local spectral 
density $\rho(x,\omega)$ near the 
end points of a 1D chain is modified compared to the bulk density. 
The algebraic behavior of the spectral density with frequency
$\omega$ close to the chemical potential was found to be governed by a
{\it boundary exponent} 
\begin{eqnarray}
\alpha_B=(K^{-1}_{\rho}-1)/z,
\label{alphaboundary}
\end{eqnarray}
which, for repulsive 
interaction ($K_{\rho} < 1$), is larger than the bulk exponent
$\alpha$. 

Systems of 1D
correlated electrons can be viewed as being at a quantum critical
point\cite{Voit}. The occurrence of nonuniversal (critical) exponents
can then be traced to the fact that the low-energy physics is governed by a
line of fixed points. From the theory of critical phenomena it is
known that critical exponents at a boundary usually differ
from their bulk counterparts. Having this in mind the observed
difference between $\alpha_B$ and $\alpha$ is not surprising\cite{VWG}. 

To investigate whether Eq.\ (\ref{alphaboundary}) also characterizes
the spectral function for models with OBC and a short range interaction 
we have calculated the spectral weight at the boundary site and the
chemical potential using the density-matrix renormalization-group
(DMRG) method\cite{dmrg} for the lattice model of spinless fermions 
with nearest neighbor interaction $U$ and the 1D Hubbard model with 
onsite interaction $U$. 
In Secs.\ \ref{spinless} and \ref{hubbard} we show that the
numerically exact data are indeed consistent with Eq.\
(\ref{alphaboundary}), although for the Hubbard model only at energies
extremely close to $\mu$. 

For small interactions and the models considered $K_{\rho}-1$ can be
expanded to give a leading behavior which is linear in the
interaction\cite{Voit,Schulz,Qin}. For $\alpha$ this gives a leading 
term which is quadratic in the 
interaction. In second order perturbation theory for the self-energy
the nonanalytic power-law behavior appears as a logarithmic
divergence $\ln{|\omega|}$ with a prefactor which is of second
order in the interaction\cite{Solyom}.
In contrast $\alpha_B$ has a contribution {\it linear} in the 
interaction. Thus signs of the nonanalytic behavior
of $\rho(x,\omega)$ can already be obtained using 
the (non-self-consistent) Hartree-Fock (HF) approximation for the 
self-energy. The least we can expect to find within the HF approximation 
is a logarithmic divergence of the above form with a prefactor linear
in the interaction. In Secs.\ \ref{TLM}, \ref{spinless}, and
\ref{hubbard} we study 
the spectral function for the three models (TL model, spinless fermions, 
Hubbard model) considered within the HF approximation. 
By comparison with the exact results (bosonization and DMRG) we show 
that many aspects of the low-energy behavior of $\rho(x,\omega)$ can be 
understood within the HF approximation. Surprisingly 
$\rho^{\rm HF}(x,\omega)$ already gives power-law
behavior. This is a very interesting observation 
as the HF approximation for the case of PBC does not capture any of
the LL features. 

Neither from general LL theory nor from the theoretical and numerical 
analysis of microscopic models, e.g.\ the 1D Hubbard model, much is known
about the energy range $\Delta$ over which the power-law behavior 
in the spectral function can be
observed. Obviously a knowledge of $\Delta$ is essential for a
meaningful comparison of theoretical and experimental spectra. 
The analytical techniques used, e.g.\ bosonization and the
Bethe Ansatz in combination with boundary conformal field
theory\cite{Voit}, only 
provide the exponent which characterizes the spectral weight at energies
{\it asymptotically} close to the chemical potential. This holds for both 
PBC and OBC. Quantum Monte-Carlo calculations of the spectral weight 
for the 1D Hubbard model with PBC 
indicate that $\Delta$ is small\cite{Preus}, which implies that
chains of many lattice sites are required to observe the suppression
in numerical calculations. 
In Sec.\ \ref{hubbard} we show that for the 
repulsive Hubbard model with OBC the spectral weight close to the 
boundary is {\it enhanced} in a large energy range around the 
chemical potential. 
The power-law suppression, predicted by bosonization, only occurs after 
a {\it crossover} at energies $\Delta$ which for small $U$ are 
{\it exponentially} (in $-1/U$) close 
to the chemical potential. For the particular case considered this
demonstrates that the bosonization (and conformal field theory) result 
only holds on a very small energy scale.   
We present an analytical discussion of the crossover behavior within 
an effective model.
From perturbation theory we expect that this kind of crossover occurs
in all models with a bare total backward scattering amplitude, i.e.\ $z$
times the $2k_F$ component of the interaction, 
which is larger than the forward scattering.

Several attempts have been made to experimentally demonstrate LL
behavior in a variety of systems which behave as
quasi one-dimensional conductors using angular integrated
and angular resolved photoemission 
spectros\-co\-py\cite{Grioni,Zwick,Denlinger,Xue,Baer}. 
Unfortunately all the measurements are plagued by various subtle 
problems\cite{Grioni} and their interpretation has been 
questioned\cite{Grioni,Xue}. 
Until three years ago the experimental spectra were 
compared to the 
predictions of bulk LL theory.
Using the results obtained from bosonization and conformal field
theory very recently many authors argued that the theoretical 
picture of a chain which, by impurities, is cut into several
disconnected pieces gives a better agreement between the
spectra observed in photoemission experiments and theoretical 
spectra\cite{EMJ,VWG,Grioni,Zwick,Baer}. 
In Sec.\ \ref{summary} we will discuss this issue
based on our results for the spectral function of the Hubbard model
and the expectation that a crossover behavior similar to the one
observed for the Hubbard model occurs also in other microscopic
lattice models. 
Our results demonstrate 
that features which occur at energies not captured by bosonization 
and conformal field theory might dominate spectra broadened by
finite temperatures
and experimental resolution. Thus a reliable interpretation of
photoemission data requires further theoretical investigations. 


\section{Open boundaries and bosonization}
\label{OBC}
In this section we present a detailed discussion of
bosonization for
a continuum model of length $L$ with OBC and a two-body 
interaction with a spatial range $R=1/q_c$.  We discuss the
subtleties which in previous approaches have only partly been 
considered\cite{FG,EMJ}.

\begin{figure}[hbt]
\epsfxsize6.0cm
\begin{center}
\epsfbox{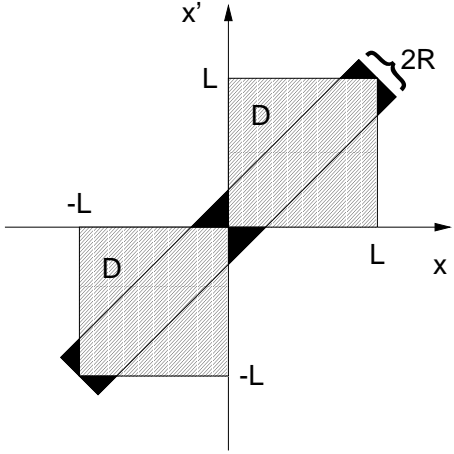}
\caption{Area of integration for the integral in Eq.\ (\ref{Fdef}). 
  For details see the text.}
\label{fig1}
\end{center}
\end{figure}

The one-particle eigenstates of 
the noninteracting system are given by
\begin{eqnarray}
\varphi_n (x) = \sqrt{2/L} \sin{ (k_n x)} ,
\label{eigenstates}
\end{eqnarray}
with $k_n = n\pi/L$,  $n \in {\bf N}$. 
Note that in contrast to the 
translational invariant case the $k_n$ do not 
have the meaning of momenta. 
A 1D system with PBC has two Fermi points $\pm k_F$.
In contrast here we only have one Fermi point given by
$k_F=n_F\pi/L$, where $z n_F$\cite{spincite} denotes the number 
of electrons in the system ($z=1$ for spinless fermions and $z=2$ for 
spin $1/2$-fermions). 
The interaction between two particles of spin species $s$ and
$s'$ is characterized by a two-body potential
$V_{s,s'}(x-x')$  
which leads to a contribution to the Hamiltonian given by
\begin{eqnarray}
\hat V & = & \frac{1}{2} \sum_{s,s'} \int^L_0 dx \int^L_0 dx'
\psi_{s}^{\dag}(x) \psi_{s'}^{\dag}(x') \nonumber \\
&& \times V_{s,s'}(x-x') \psi_{s'}^{}(x') \psi_{s}^{}(x) 
\nonumber \\
& = & \frac{1}{2} \sum_{s,s'} \int^L_0 dx \int^L_0 dx'
\hat{\rho}_{s}(x) 
V_{s,s'}(x-x') \hat{\rho}_{s'}(x')  \nonumber \\ 
&& - \frac{1}{2} \sum_{s} V_{s,s}(0) \hat{N}_{s},
\label{interactionI}
\end{eqnarray}
with the field operators $\psi_{s}^{(\dag)}(x) = \sum_{n=1}^{\infty}
\varphi_n (x) a^{(\dagger)}_{n,s} $, the density operators 
$\hat{\rho}_{s}(x)= \psi_{s}^{\dag}(x) \psi_{s}^{}(x)$,
and the particle number
operators $\hat{N}_{s} =  \int^L_0 dx \hat{\rho}_{s}(x)$.  
We express the interaction
in terms of the creation and annihilation operators
$a^{(\dagger)}_{n,s}$ of
the eigenstates $\varphi_n$ with spin $s$. 
After rearranging the terms in a way which simplifies the bosonization
discussed later it reads
\begin{eqnarray}
\hat V & = &\frac{1}{2}\sum_{s,s'} 
\left[
 \sum_{n \neq n'} \sum_{m \neq m'} v_{nmn'm'}^{s,s'} \, a^\dagger_{n,s} 
a^{}_{n',s}  a^\dagger_{m,s'} a^{}_{m',s'} \right. \nonumber \\
&& + \sum_{n \neq n'} \sum_{m} v_{nmn'm}^{s,s'} \left( a^\dagger_{n,s} 
a^{}_{n',s}  n_{m,s'} + \mbox{h.c.} \right) \nonumber \\
&&\left. + \sum_{n,m} v_{nmnm}^{s,s'} n_{n,s} n_{m,s'}
- \delta_{s,s'} V_{s,s}(0) \hat{N}_{s} \right],
\label{interactionII}
\end{eqnarray}
with the matrix elements 
\begin{eqnarray}
v_{nmn'm'}^{s,s'} & = & 
\int^L_0 dx \int^L_0 dx' \varphi^{\ast}_{n}(x) 
\varphi^{}_{n'}(x) \nonumber \\ && \times V_{s,s'}(x-x') 
\varphi^{\ast}_{m}(x') \varphi^{}_{m'}(x') 
\label{matrixelementI}
\end{eqnarray}
and the occupation number operators $n_{n,s}= a^\dagger_{n,s} a^{}_{n,s}$.
If we express
products of the sine functions $\varphi_n$
in terms of cosine functions they read
\begin{eqnarray}
&& v_{nmn'm'}^{s,s'} = 
\left[ F_{s,s'} \left(k_n-k_{n'},k_m-k_{m'}\right) \right.
\nonumber \\
&& - F_{s,s'}\left(k_n-k_{n'},k_m + k_{m'}\right)  
- F_{s,s'} \left(k_n+k_{n'},k_m - k_{m'}\right) \nonumber \\ 
&& \left. + F_{s,s'}\left(k_n + k_{n'},k_m + k_{m'}\right) \right]/L,
\label{matrixelementII}
\end{eqnarray}
with
\begin{eqnarray}
F_{s,s'}(q,q')  & = &
\frac{1}{L} 
\int^L_0 dx \int^L_0 dx' \cos{( qx)} V_{s,s'}(x-x')   \nonumber \\ 
&& \times \cos{(q'x')} \nonumber \\
 & = & \frac{1}{4L} \int \!\!\! \int_{D} dx \,
  dx' \left\{ e^{i (qx -q'x')} + e^{i (qx + q'x')} \right\} \nonumber
  \\ 
&& \times V_{s,s'}(x-x').  
\label{Fdef}
\end{eqnarray}
The area $D$ over which we have to integrate in 
Eq.\ (\ref{Fdef}) is given by the two
hatched squares shown in Fig.\ \ref{fig1} each of size
$L^2$. 
Using the fact that $V_{s,s'}(x)$ was assumed to have a range $R$, 
$F_{s,s'}(q,q')$ can partly be expressed in terms of the 
Fourier transform $\tilde{V}_{s,s'}(q)=\int_{-L}^{L} dx
V_{s,s'}(x) \exp{(-iqx)}$ if we instead 
integrate over the rectangle 
of width $2R$ indicated in Fig.\ \ref{fig1}. This can be achieved 
by adding and subsequently subtracting the integral over the six black 
triangles. 
If we denote their contribution by $-g_{s,s'}(q,q')/L$, we obtain
\begin{equation}
F_{s,s'}(q,q') =  \tilde{V}_{s,s'}(q)/2 \, 
\left( \delta_{q,q'} +
  \delta_{q,-q'}\right) + g_{s,s'}(q,q')/L.
\label{Fcalculated}
\end{equation}
In the case of a spin independent exponential interaction $V(x)=V_0/(2R)
\exp{(-|x|/R)}$, $g(q,q')$ is proportional to
$\tilde{V}(q)\tilde{V}(q') /V_0$. 

For comparison we give a brief discussion of the interaction part of
the Hamiltonian for the 1D  electron gas with PBC.
The noninteracting eigenstates are
plane waves $\phi_{n}(x) = 
\exp{(i\tilde{k}_{n} x)}/\sqrt{L}$ with $\tilde{k}_{n}
= 2 \pi n /L$ and $n \in {\bf Z}$. In terms of
the creation and annihilation operators $\tilde{a}_{n,s}^{(\dag)}$ 
of these states a two-particle interaction gives a term similar to
Eq.\ (\ref{interactionII}) with the matrix elements 
\begin{eqnarray}
\tilde{v}_{nmn'm'}^{s,s'}  =
\tilde{F}_{s,s'}(\tilde{k}_n-\tilde{k}_{n'}, 
\tilde{k}_m-\tilde{k}_{m'}) 
\label{matrixelementpbcI}
\end{eqnarray}
and 
\begin{eqnarray}
\tilde{F}_{s,s'}(q,q') = \tilde{V}_{s,s'}(q)
\, \delta_{q,-q'} / 2 .
\label{matrixelementpbcII}
\end{eqnarray} 
For PBC the matrix elements thus only depend on the momentum transfer
$\tilde{k}_n-\tilde{k}_{n'} = \tilde{k}_{m'} - \tilde{k}_m$ between
the two scattering particles. The part of the interaction with small
momentum transfer $q \ll k_F$ can be written as a bilinear form in
the spin and charge density operators of left and right moving 
fermions\cite{Tomonaga,Luttinger,Bosintr}. 
There are two ways of treating the 
backscattering processes with momentum
transfer $2 k_F$. One can show that for repulsive interactions
backscattering is irrelevant in the RG sense\cite{Solyom}. 
If one is only interested in the ``critical'' low-energy properties of
the model one can thus neglect backscattering from the beginning. Then 
the physical two-body potential has to be replaced by effective
coupling constants, as it is changed by the initial flow of the 
backscattering and one ends up with an effective low-energy 
model\cite{Solyom}. 
The other way to proceed is to assume that the interaction 
$\tilde{V}_{s,s'}(q)$ is cut off at a momentum $q_c=1/R \ll k_F$, 
i.e.\ is long range in real space and only consider a subspace of the
Fock space ${\mathcal F}_T$ with no holes deep in the Fermi see and 
no particles in 
one-particle states with energies much higher then the Fermi energy.
This is the idea originally considered by Tomonaga\cite{Tomonaga}. 
On this subspace the properly normalized density operators of left 
and right moving fermions obey bosonic commutation
relations\cite{Bosintr}. For a one-particle dispersion which is
linearized around the two Fermi points also the kinetic energy is
quadratic in the densities\cite{Bosintr} and the entire Hamiltonian   
can be written as a Hamiltonian of noninteracting bosons. 

From Eqs.\ (\ref{matrixelementII}) and (\ref{Fcalculated}) it is
obvious that the situation is more complex for the case of OBC. 
Expressed in the basis of the noninteracting eigenstates the
interaction contains a variety of scattering processes. 
Only a few of them can be
written bilinearly in the density operators ($m \in {\bf Z}$, $m
\neq 0$)  
\begin{equation}
\rho_{m,s}= 
\sum_{n=\mbox{\scriptsize max}\{1,1-m\}}^{\infty}
a_{n,s}^{\dag} a_{n+m,s}^{},
\label{bosonendef}
\end{equation}
which obey $\rho_{m,s}^{\dag}= \rho_{-m,s}^{}$.
Alternatively charge and spin density operators
\begin{eqnarray}
\rho_{m,\rho} & = & (\rho_{m,\uparrow} +
\rho_{m,\downarrow})/\sqrt{2} \\
\rho_{m,\sigma} & = & (\rho_{m,\uparrow} -
\rho_{m,\downarrow})/\sqrt{2} 
\label{bosonendefcs}
\end{eqnarray}
can be used. 
These operators are defined in analogy to the case of PBC.
A RG study which includes all the scattering vertices is still 
missing and we will thus follow the idea of Tomonaga to investigate
whether this leads to a Hamiltonian bilinear in the densities similar
to PBC.  Thus we assume that the interaction $V_{s,s'}(x)$
is long range in real space and only consider the low-energy 
subspace ${\mathcal  F}_T$.  
For simplicity we will furthermore focus
on the case of a
spin independent interaction $V_{s,s'}(x)= V(x)$.  
Acting on states $\left| \psi_T \right> \in {\mathcal  F}_T$ the
interaction term $\hat V$ simplifies because $a_{n,s}^{\dag} a_{n',s}
\left| \psi_T \right> = 0$ for $0< |n-n'| \leq n_c$, 
unless $k_{n'}$ is close
to $k_F$. As, with our assumption about the range of the interaction,
$F(q,q')$ is only nonzero if both arguments are smaller than $q_c$ we
can replace $v_{nmn'm'}$ in the first term on the right hand side of
Eq.\ (\ref{interactionII}) by $F(k_n - k_{n'}, k_m - k_{m'})/L$. With
similar arguments for the other contributions the interaction term on
${\mathcal  F}_T$ reads
\begin{eqnarray}
\hat V & = &\frac{1}{2L} 
{\sum_{n,n'}}^{\prime}
 F_{\rho}(q_n, q_{n'})  
\rho_{n,\rho}^{} \rho_{n',\rho}^{} \nonumber \\
&&+\frac{1}{L}  \sum_{n>0}  F_{\rho}(q_n, 0)
 \hat{N}_{\rho}  \left( \rho_{n,\rho}^{} +
   \rho_{-n,\rho}^{}\right)  \nonumber \\
&& - \frac{\sqrt{z}}{L}  \sum_{n>0} 
  \sum_{m>0} F_{\rho}(q_n, 2 k_m) \left( \rho_{n,\rho}^{} +
   \rho_{-n,\rho}^{}\right) \nonumber \\
&&+ \frac{1}{2L} F_{\rho}(0, 0) \hat{N}_{\rho}^2 - \frac{\sqrt{z}}{L}
\sum_{m>0} F_{\rho}(0, 2 k_m)  \hat{N}_{\rho}
\nonumber \\ 
&& - \frac{\sqrt{z}}{2} \, V_{\rho}(0) \hat{N}_{\rho} 
+\frac{1}{L} \sum_{m,m'>0}  F_{\rho}(2 k_m, 2 k_{m'}) ,
\label{interactionIII}
\end{eqnarray}
with 
\begin{eqnarray}
\hat{N}_{\rho} & = & (\hat{N}_{\uparrow} +
\hat{N}_{\downarrow})/\sqrt{2}, \\  
\label{Nrhodef}
F_{\rho}(q, q')  & = & z F_{s,s'}(q, q'), \\
\label{Frhodef}
V_{\rho}(0) & = &   z V_{s,s'}(0) .
\label{Vrhodef}
\end{eqnarray}
The prime at the sum in Eq.\ (\ref{interactionIII}) indicates that
the terms with $n=0$ or $n'=0$ are excluded. For $z=1$ the above 
formulas give the
corresponding expressions for the case of spinless fermions, if one
puts $\rho_{n,\rho} \to \rho_n$ and $\hat{N}_{\rho} \to \hat{N}$.  
Compared to the analogous expression for PBC\cite{Bosintr} 
Eq.\ (\ref{interactionIII}) contains three modifications: 
Generically $F_{\rho}(q_{n},q_{n'})$ is nonvanishing for all
$q_{n}$ and $q_{n'}$ smaller than $q_c$   
which leads to a coupling between all the $\rho_{n,\rho}$ in the first
line of Eq.\ (\ref{interactionIII})\cite{FG}. 
For PBC only $\rho_{n,\rho}$ with $n=-n'$ are coupled.
OBC furthermore lead to 
a term which couples the particle number and the density operators
and a term linear in the  $\rho_{n,\rho}$\cite{Tobias}. 
As we will discuss in Sec.\ \ref{HF} the
latter leads to a nontrivial contribution to the Hartree term in the
self-energy. 
For the discussion of the one-particle properties the constant as well
as terms linear in $\hat{N}_{\rho}$ can be neglected as the latter
only lead to a renormalization of the chemical
potential\cite{Bosintr}. If we linearize the dispersion around $k_F$
and for $n>0$ define $b_{n,\nu}= \rho_{n,\nu}/\sqrt{n}$ for $\nu=\rho,
\sigma$ the kinetic
energy on ${\mathcal F}_T$ can, up to particle number contributions, be 
replaced by
\begin{eqnarray}
\hat{H}_0 = v_F \frac{\pi}{L} \sum_{n>0} n \left( 
b_{n,\rho}^{\dag} b_{n,\rho}^{} + b_{n,\sigma}^{\dag} b_{n,\sigma}^{}
\right)
\label{Tdef}
\end{eqnarray}
with the Fermi velocity $v_F$. 
On ${\mathcal  F}_T$ the 
$b_{n,\rho/\sigma}^{}$ obey bosonic commutation relations and  
charge and spin degrees of freedom do commute.  
Note that in $\hat{H} = \hat{H}_0 + \hat {V}$ similar to PBC spin and 
charge degrees of freedom are decoupled. 
Formally a
Hamiltonian $\hat{H}$ of coupled and shifted
harmonic oscillators can be diagonalized using a Bogoljubov
transformation. For a general interaction the transformation cannot be
given analytically, but e.g.\ for the above mentioned interaction 
$V(x)=V_0/(2R) \exp{(-|x|/R)}$, exploiting the fact that the
corresponding $g(q,q')$ is separable, it can
analytically be shown that neither the coupling 
of bosons with $\left|q_{n}\right| \neq \left|q_{n'}\right|$ given by
$g(q_n,q_{n'})$ nor the terms linear in the bosons do change the
``critical'' one-particle properties of the model, i.e.\ the 
exponents of the asymptotic decay of correlation functions\cite{Tobias}. 
As long as we are only
interested in the ``critical'' behavior we can thus work with the
simplified Hamiltonian
\begin{eqnarray}
\hat{H} & = &  \sum_{n>0} n \left[ v_F \frac{\pi}{L}  
b_{n,\rho}^{\dag} b_{n,\rho}^{} + \frac{z}{4 L} \tilde{V}(q_n) 
\left( b_{n,\rho}^{\dag} + b_{n,\rho}^{} \right)^2 \right]
\nonumber \\ && + v_F \frac{\pi}{L}   
\sum_{n>0} n b_{n,\sigma}^{\dag} b_{n,\sigma}^{} .
\label{TLsimpledef}
\end{eqnarray}
In the following we will denote the model related to this Hamiltonian
the TL model, even though we have dropped  terms 
compared to Eq.\ (\ref{interactionIII}) as described above.
To obtain the spinless model from Eq.\ (\ref{TLsimpledef}) 
one has to drop the second line and set $z=1$.
The TL model with OBC consists of an independent system of selfcoupled
oscillators. This is in contrast to the PBC case, where the modes $n$
and $-n$  are coupled by the interaction term.
Apart from a constant Eq.\ (\ref{TLsimpledef}) can be brought 
into the form\cite{FG,Bosintr}
\begin{eqnarray}
\hat{H}  = \sum_{n>0} \left[ \omega_n \alpha_{n,\rho}^{\dag} 
\alpha_{n,\rho}^{} +v_F \frac{\pi}{L} n b_{n,\sigma}^{\dag}
b_{n,\sigma}^{} \right] ,
\label{TLsimplediag}
\end{eqnarray}
with $\omega_n = k_n \sqrt{1+ z \tilde{V}(k_n)/(\pi v_F)}$ and bosonic
operators $\alpha_{n,\rho}$ given by a linear combination of
$b_{n,\rho}^{\dag}$ and $b_{n,\rho}^{}$.  

The one-particle spectral function 
$\rho^<(x,\omega)$ relevant for photoemission 
can be determined from the Green's function
$G^<(x,x,t)=  -i \left< \psi^{\dag}_s (x,0) \psi^{}_s (x,t) \right> $
using\cite{spinindependent}  
\begin{eqnarray}
\rho^<(x,\omega) = \int_{-\infty}^{\infty} \frac{dt}{2 \pi} e^{i \omega
  t} i G^<(x,x,t) .
\label{Greendef}
\end{eqnarray}
Here $\left<  \ldots \right> $ denotes the ground state expectation
value. For the TL model with OBC $G^<(x,x,t)$ can be calculated using the
bosonization of the fermionic field operator 
\begin{eqnarray}
\psi_{s}^{}(x) & = &\sqrt{\frac{2}{L}} \sum_{n=1}^{\infty}
\sin{(k_n x)} \,  a^{}_{n,s} \nonumber \\
& = & \frac{-i}{\sqrt{2 L}} \sum_{n=1}^{\infty} 
\left[ e^{i k_n x} -   e^{-i k_n x}   \right]a^{}_{n,s}.  
\label{psidef}
\end{eqnarray}
Eq.\ (\ref{psidef}) cannot directly be expressed in the bosons
$b_{n,\rho}^{(\dag)}$  and $b_{n,\sigma}^{(\dag)}$. To achieve this we
have to add one-particle states with quantum numbers $k_n$, $n<1$ to
the Hilbert space. They are assumed to be filled in the ground state
and do not modify the low-energy physics of the model. The auxiliary
field operator 
\begin{eqnarray}  
\tilde{\psi}_{s}^{}(x) = \frac{1}{\sqrt{ L}} \sum_{n=-\infty}^{\infty} 
e^{i k_n x} a^{}_{n,s}  
\label{psiauxdef}
\end{eqnarray}
can then be written in the boson operators similar to
PBC\cite{Bosintr} and $G^<(x,x,t)$ is given by
\begin{eqnarray}  
&& G^<(x,x,t)  =  \left[  
\left< \tilde{\psi}^{\dag}_s(x,0) \tilde{\psi}^{}_s(x,t) \right> 
+ \left< \tilde{\psi}^{\dag}_s(-x,0) \tilde{\psi}^{}_s(-x,t) \right>
\right. \nonumber \\
&& - \left. 
\left< \tilde{\psi}^{\dag}_s(x,0) \tilde{\psi}^{}_s(-x,t) \right> 
- \left< \tilde{\psi}^{\dag}_s(-x,0) \tilde{\psi}^{}_s(x,t) \right> 
\right]/2 .
\label{Greenaux}
\end{eqnarray}
The expectation values in Eq.\ (\ref{Greenaux}) can be calculated
using bosonization of the auxiliary field operator\cite{Bosintr}. 
For a fixed
position $x$ the leading behavior of $G^<(x,x,t)$ at large times 
is\cite{FG} 
\begin{eqnarray}  
G^<(x,x,t)  \sim t^{-(1+\alpha_B)} 
\label{tasymp}
\end{eqnarray}
with 
\begin{eqnarray}  
\alpha_B = (K^{-1}_{\rho}-1)/z
\label{alphaBdef}
\end{eqnarray}
and the LL parameter
\begin{eqnarray}   
K_{\rho} = \left[ 1 + \frac{z\tilde{V}(0)}{\pi v_F}\right]^{-1/2} 
\label{LLparameterTL}
\end{eqnarray}
of the TL model.
The nonanalytic behavior of the spectral function close to the
chemical potential is thus given by 
\begin{eqnarray} 
\rho^<(x,\omega) \sim |\omega|^{\alpha_B} \Theta(-\omega) . 
\label{omasymp}
\end{eqnarray}

Without explicitly demonstrating that the TL model is the effective
low-energy model (fixed point model) for all models of LL's with OBC the
results of Eqs.\ (\ref{alphaBdef}) and (\ref{omasymp}) have been
assumed to hold for all LL's\cite{FG,EMJ}. This implies that the boundary 
exponent $\alpha_B$ can be expressed in terms of the bulk LL parameter
$K_{\rho}$. The generalization found some confirmation in 
Ref.\ \cite{WVP} where methods of boundary conformal field 
theory were used to calculate $\alpha_B$ for Bethe ansatz solvable
models. In Sec.\ \ref{spinless} we will present
numerically exact results for the spectral weight of the lattice
model of spinless fermions with nearest neighbor interaction, which
are consistent with Eq.\ (\ref{omasymp}) over a fairly large energy
range. 
Our results for the Hubbard model presented in Sec.\ \ref{hubbard} 
demonstrate that in this model the spectral function displays a richer
structure. Only for energies $\Delta$ exponentially (in $-1/U$) 
close to $\mu$, i.e.\ exponentially large system sizes, the small $U$ 
data are consistent with  Eq.\ (\ref{omasymp}). 

Within the TL model and for fixed interaction strength 
the $x$ dependence of the energy range $\Delta$ 
over which the asymptotic power-law suppression with the boundary 
exponent $\alpha_B$ can be observed is given by \linebreak 
$\Delta \approx v_F \, \mbox{min}\{ 1/x, 1/R \}$. For $x>R$ and $v_F/x <
\omega < v_F/R$  
the bulk ``critical'' behavior with a power-law suppression
given by $\alpha$ defined in Eq.\ (\ref{alphabulk}) is recovered.
For $\omega > v_F/R$ ``nonuniversal'' features dominate the
spectrum. 
Unfortunately the energy scales obtained from the TL model 
cannot be trusted if one is interested in  more 
realistic continuum or lattice models with a nonlinear one-particle 
dispersion and scattering processes given by a general interaction
(not necessarily long range in real space).
The best bosonization can provide for such models is the exponent of the
nonanalytic behavior at energies asymptotically close to $\mu$. 

For small $\tilde{V}(0)$ it follows from Eq.\ (\ref{LLparameterTL}) that 
\begin{eqnarray}
K_{\rho} = 1- \frac{z\tilde{V}(0)}{2 \pi v_F} + {\mathcal O}\left( \left[ 
\frac{\tilde{V}(0)}{\pi v_F} \right]^2\right)
\label{Krhoexpan}
\end{eqnarray}
and thus  from Eq.\ (\ref{alphaBdef}) that
\begin{eqnarray}  
\alpha_B = \frac{\tilde{V}(0)}{2 \pi v_F} + {\mathcal O}\left( \left[ 
\frac{\tilde{V}(0)}{\pi v_F} \right]^2\right)
\label{alphaBexpan}
\end{eqnarray}
This has to be contrasted to the small $\tilde{V}(0)$ behavior of the
bulk exponent $\alpha$ Eq.\ (\ref{alphabulk}) which is quadratic in
the interaction. Thus signs of the nonanalytic behavior
of $\rho(x,\omega)$ can already be obtained using 
the HF self-energy, which will be analysed in detail in the 
next section.


\section{Hartree-Fock self-energy}
\label{HF}

In this section we discuss the non-self-consistent HF approximation
for the  self-energy $(\Sigma^{\rm HF}_s)_{n,n'}$,  for 
the continuum  model
defined by Eq.\ (\ref{interactionI}) and a one-particle dispersion 
$\varepsilon(k)$. In contrast to the end of the
last section we do not restrict our discussion to the subspace 
${\mathcal F}_T$ and thus consider all the scattering processes given
by the Eqs.\ (\ref{interactionII}) to (\ref{Fcalculated}).
We assume  that the  interaction is spin independent. For PBC 
$(\Sigma^{\rm HF}_s)_{n,n'} \propto \delta_{n,n'}$ because of momentum
conservation and the HF approximation only leads to finite shifts 
in $\mu$ and $v_F$. It does not capture any of the peculiar properties
of LL's.  
As already mentioned following Eq.\ (\ref{Tdef}) $g_{s,s'}(q,q')/L$ in Eq.\
(\ref{Fcalculated}) does not contribute to the ``critical'' LL 
properties of the continuum model with OBC. 
Later we will confirm this using perturbation theory.
We will thus neglect this 
term and only consider the $\delta$ terms in Eq.\ (\ref{Fcalculated}). 
With the matrix elements given by Eq.\ (\ref{matrixelementII}) this 
leads to 
\begin{eqnarray}
&& \left[ \Sigma^{\rm HF}_s \right]_{n,n'}  = \sum_{n''=1}^{n_F}
\sum_{s'} \left(  v_{n n'' n' n''}^{s,s'}  - v_{n n'' n'' n'}^{s,s'}
 \delta_{s,s'}\right) \label{selfenergydef} \\ 
&& = \delta_{n,n'} \left\{ \delta \mu-
 \frac{1}{2 L} \sum_{n_1 = 1}^{n_F} \left[ \tilde V(k_n-k_{n_1})  
+ \tilde V(k_n+k_{n_1}) \right]  \right\} \nonumber\\ 
 && + 
 \frac{1}{2 L} \left\{ z\tilde V(k_n+k_{n'})- \tilde V
  \left( \frac{k_n-k_{n'}}{2} \right) \right\}
f\left( \frac{n+n'}{2} \right) \nonumber  \\
&& - 
 \frac{1}{2 L} \left\{ z \tilde V(k_n-k_{n'}) - \tilde V
  \left(\frac{k_{n}+k_{n'}}{2} \right) \right\} 
f\left(\frac{|n-n'|}{2}\right) \nonumber ,
\end{eqnarray}
with 
\begin{eqnarray}
f(m) = \left\{
            \begin{array}{cl}
            1  & \mbox{for} \,\, m \in \{1,2,\ldots, n_F \}  \\
           0 & \mbox{otherwise}
\end{array} \right.
\label{kleinfdef}
\end{eqnarray}
and $\delta \mu =z\tilde V(0) n_F/L$. Due to the parity symmetry with
respect to the middle of the box only matrix elements with even $n \pm n'$
are nonvanishing. Note that the self-energy is
$\omega$ independent and real but has a nontrivial matrix structure
in the quantum numbers $n$ and $n'$ due to the broken translational 
invariance. From the self-energy the retarded Green's function 
follows by a matrix inversion\cite{spinindependent}
\begin{eqnarray}
\left[G(\omega)\right]_{n,n'} = 
\left[ \left\{ \omega - \xi(k_n) + i0 \right\} {\mathbf 1}  - 
\Sigma \right]_{n,n'}^{-1} ,
\label{Gretard}
\end{eqnarray} 
with the unity matrix ${\mathbf 1}$ and $\xi(k) = \varepsilon(k) -
\mu$.  
The local spectral function is then given by
\begin{eqnarray}
\rho(x,\omega) = - \frac{1}{\pi} \, \mbox{Im} \, 
\sum_{n,n'=1}^{\infty} \varphi_n^{\ast}(x)
\varphi_{n'}^{}(x) 
\left[G(\omega)\right]_{n,n'} .
\label{specdef}
\end{eqnarray}    
$\rho^{<}(x,\omega)$ defined in Eq.\ (\ref{Greendef}) is related 
to the total spectral function of Eq.\ (\ref{specdef}) by
$\rho^{<}(x,\omega) = \rho(x,\omega) \Theta(\omega)$.  
To specify the model we have to chose a dispersion
$\varepsilon(k)$ which is assumed to be sufficiently smooth. 
In the thermodynamic limit the noninteracting spectral weight is  
given by\cite{disp} 
\begin{eqnarray}
\rho^0(x,\omega) = \left. \frac{1}{\pi} \left| \frac{d \varepsilon(k)}{dk}
\right|^{-1} \left[ 1-\cos{\left(2 k x \right) }  \right]
\right|_{k=\xi^{-1}(\omega)} , 
\label{specnoint}
\end{eqnarray}     
where $\xi^{-1}(\omega)$ denotes the function inverting $\xi(k)$.  
We did not succeed in analytically inverting the matrix 
$\left[ G^{\rm HF}(\omega) \right]^{-1}$ in Eq.\ (\ref{Gretard})
which, for a general interaction, has a nontrivial structure. 
To gain a first insight in the behavior of the spectral weight close to
$\mu$ the matrix can be inverted perturbatively to lowest order 
in $\tilde{V}$
\begin{eqnarray}
\rho^{\rm HF}(x,\omega) & = &\rho^{0}(x,\omega) \left\{ 1 + 
  \left[ \frac{\tilde{V} (0)}{2 \pi v_F} - z \frac{\tilde{V}(2
      k_F)}{2 \pi v_F} \right] \right. \nonumber \\
&& \left. \times  \ln  \left|\frac{\omega}{v_F k_F} \right|
 + {\mathcal O}\left(\tilde{V}^2\right)
 \right\} .
\label{specpert}
\end{eqnarray}  
Note that the logarithmic divergence  in $\rho^{\rm HF}(x,\omega)$
is not due to a singular 
frequency behavior of $\Sigma^{\rm HF}$, but emerges in the 
perturbative approach to the 
matrix inversion. This has to be contrasted to PBC where 
the first indication of a break down of perturbation theory can be 
found in second order. The second order self-energy $\Sigma^{(2)}$
displays a 
logarithmic divergence $\tilde{V}^2 \ln{|\omega|}$ leading to the 
same kind of divergence in the spectral
function\cite{Solyom,Bosintr}. 
The diagram responsible for the second order divergence for 
PBC does also lead to a logarithmic divergence 
\begin{eqnarray}
\left[ \Sigma^{(2)}_s(\omega) \right]_{n,n'} 
\sim \delta_{n,n'}\tilde{V}^2 
\ln{|\omega|} 
\label{obcsecondorder}
\end{eqnarray}
for OBC. Thus {\it logarithmic} terms found in perturbation theory for 
$\rho(x,\omega)$ in the case of OBC can have {\it two different origins.} 

The leading
logarithmic behavior in  $\rho^{\rm HF}(x,\omega)$ Eq.\ (\ref{specpert})
comes from the
term in $\Sigma^{\rm HF}$ which is proportional to
$f([n+n']/2)$. It gives a sharp step in the self-energy
matrix [see Eq.\ (\ref{kleinfdef})] which crosses the diagonal 
at $(n_F,n_F)$. The height of the step determines the prefactor of the
logarithmic divergence. Contributions of $g(q,q')$ neglected in Eq.\
(\ref{selfenergydef}) are continous 
and thus do not modify the nonanalytic behavior,
consistent with the observation made in the last section.
In Sec.\ \ref{hubbard} we will calculate the HF self-energy for the
Hubbard model with OBC and present an analytical discussion of the
resulting spectral weight in an effective model. 
Sharp steps in the HF self-energy matrix will also play a prominent 
role in this discussion.

To investigate how the leading 
logarithmic divergence of $\rho^{\rm HF}(x,\omega)$ 
Eq.\ (\ref{specpert})
is modified by higher orders in $\tilde{V}$, 
we numerically invert 
$\left[ G^{\rm HF}(\omega) \right]^{-1}$ for different models 
of finite size 
in Secs.\ \ref{TLM}, \ref{spinless}, and \ref{hubbard}. 
It turns out that already the HF approximation leads to a power-law. 
We will
furthermore compare the spectral weight in the HF approximation to
exact results obtained from bosonization and DMRG.


\section{Tomonaga-Luttinger model}
\label{TLM}

\begin{figure}[thb]
\begin{center}
\epsfxsize6.7cm
\epsfbox{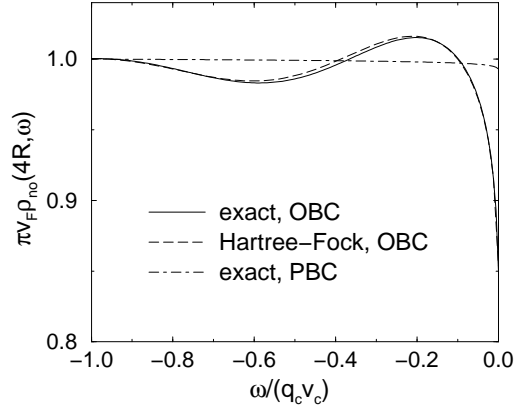}
\caption{Spectral weight near the boundary ($x=4R$)
for the spinless TL model with $U/(2\pi v_F) =0.05263$, 
$n_c \equiv q_c L / \pi=160$, and $n_F = 4 n_c$.}
\label{fig2}
\end{center}
\end{figure}

The spectral function for the TL model of finite length $L$ 
defined by Eq.\ (\ref{TLsimpledef}) can be calculated exactly using
bosonization and a recursive method introduced in Ref.\ 
\cite{KSVM} for the case of PBC. For the momentum integrated
spectral function of the TL model the spin only leads to factors of
two and we will thus consider the spinless TL model.
The last two terms in Eq.\
(\ref{Greenaux}) give contributions to  $\rho(x,\omega)$ which
for energies close to the chemical potential are proportional to
$\cos{(2 k_F x)}$. Such a term is already present in the
noninteracting spectral function Eq.\ (\ref{specnoint}). It drops out
if the spectral function is spatially averaged over a small
length. 
If one is interested in
the comparison to experimental spectra this assumption is justified
as photoemission measurements will automatically average over
some small spatial range.
In Fig.\ \ref{fig2} only the nonoscillatory part $\rho_{\rm no}(x,\omega)$
of $\rho(x,\omega)$ is shown (solid line) as a function of $\omega$.
The low-energy behavior is independent of the shape of the interaction 
and we have chosen the simple form $\tilde{V}(q) = U \Theta(q_c-|q|)$.
For a finite system $\rho(x,\omega)$ is given by a sum of $\delta$
peaks. In Fig.\ \ref{fig2} the energies with 
nonvanishing weight are equidistant 
with a level spacing $\propto 1/L$ 
and we have connected them to a continuous line. 
A power-law behavior of the weight close to $\mu$ can only be found 
in the thermodynamic limit, but for
large system sizes the weight of the $\delta$ peaks resembles the 
power-law obtained for $L \to \infty$.
The data in Fig\ \ref{fig2} display the
suppression of spectral weight close to $\mu$ which is much stronger
then the one observed for PBC. For comparison PBC data for the same
parameters are shown in Fig.\ \ref{fig2} as a dashed-dotted line.

\begin{figure}[thb]
\begin{center}
\epsfxsize4.0cm
\epsfbox{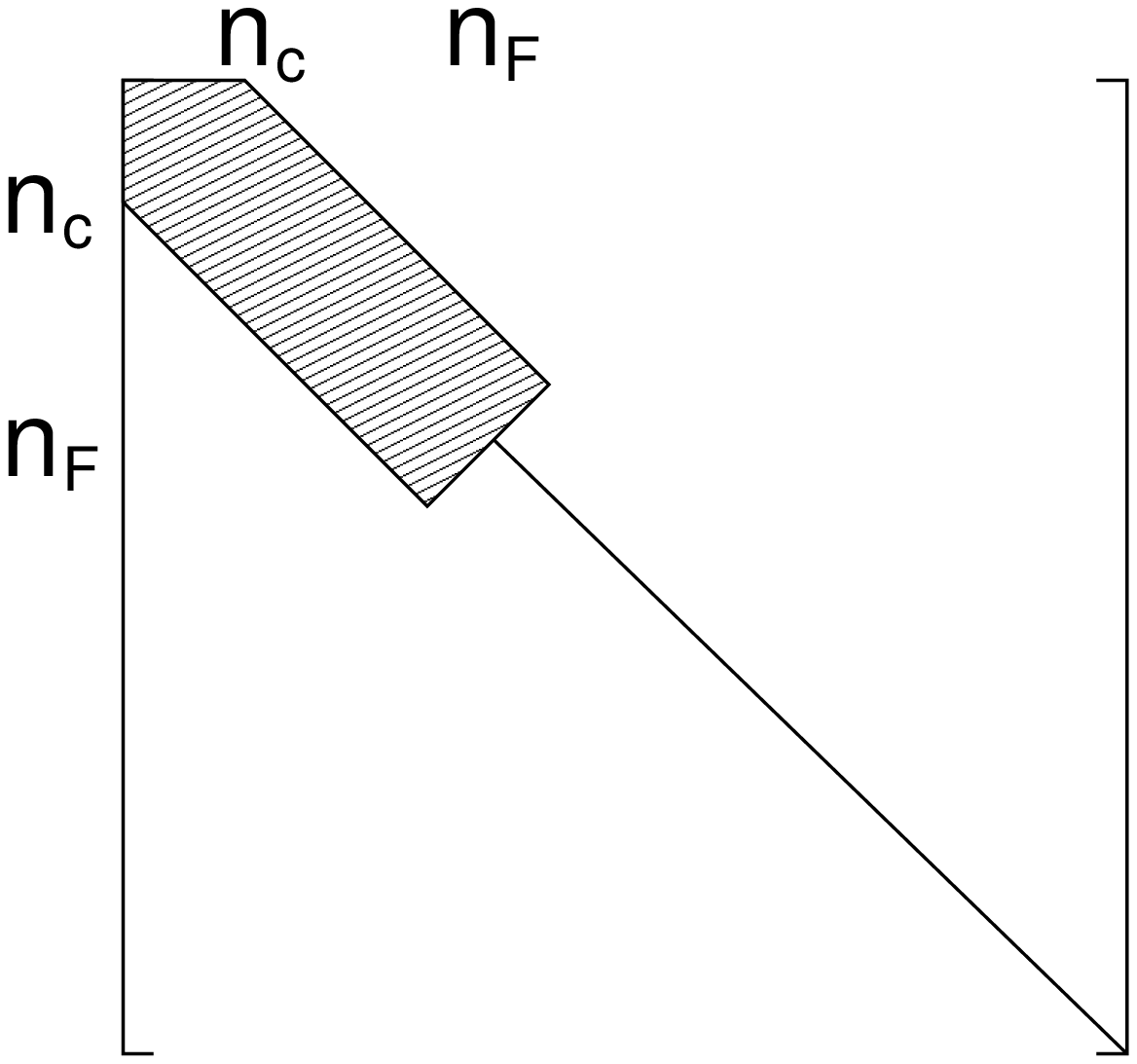}
\caption{$\left[ G^{\rm HF} \right]^{-1}$ for the TL model. For
  details see the text.}
\label{fig3}
\end{center}
\end{figure}

The HF self-energy for the TL model differs for the two choices Eqs.\
(\ref{interactionIII}) and (\ref{TLsimpledef}) for the interaction. 
The term proportional to $\tilde{V}(k_n-k_{n'}) f(|n-n'|/2)$ in 
Eq.\ (\ref{selfenergydef}) is due to the
interaction term linear in the bosons neglected in 
Eq.\ (\ref{TLsimpledef}). As
it does not contribute to the singular part of the HF spectral  
function we treat the HF approximation to the TL model 
Eq.\ (\ref{TLsimpledef}). This leads to
\begin{eqnarray}
\left[ \Sigma^{\rm HF} \right]_{n,n'} & = & 
\delta_{n,n'} \left\{ z \frac{n_F}{L} \tilde{V}(0) -
 \frac{1}{2 L} \sum_{n_1 = 1}^{n_F}  \tilde V(k_n-k_{n_1})  
 \right\} \nonumber\\ 
&& -  \frac{1}{2 L}  \tilde V \left( \frac{k_n-k_{n'}}{2} \right)
f\left( \frac{n+n'}{2} \right) .
\label{selfenerTL}
\end{eqnarray}
The first term in Eq.\ (\ref{selfenerTL}) is the HF shift of
the chemical potential and the second term leads to a renormalized  Fermi
velocity $v_{\rm HF} = v_F + \tilde{V}(0)/ (2 \pi)$. The resulting
matrix $\left[ G^{\rm HF} \right]^{-1}$ which has to be
inverted is sketched in Fig.\ \ref{fig3}. If we consider the box
shaped potential $\tilde{V}(q) = U \Theta(q_c-|q|) $
the last term in Eq.\ (\ref{selfenerTL}) does contribute as a constant
in the hatched area. 
In Fig.\ \ref{fig2}
$\rho^{\rm HF}_{\rm no}(x,\omega)$ is shown as the dashed
line. For the parameters chosen (weak interaction)  
the HF result agrees quantitatively with the exact solution.  
For the TL model $\tilde{V}(2 k_F) = 0$, which leads to 
[see Eq.\ (\ref{specpert})] 
\begin{eqnarray}
\rho^{\rm HF}(x,\omega)  \sim   1 + 
  \frac{\tilde{V} (0)}{2 \pi v_F} \, \ln 
   \left|\frac{\omega}{v_F k_F} \right| +  \ldots \,\, .  
\label{specpertTL}
\end{eqnarray}

\begin{figure}[thb]
\begin{center}
\epsfxsize6.7cm
\epsfbox{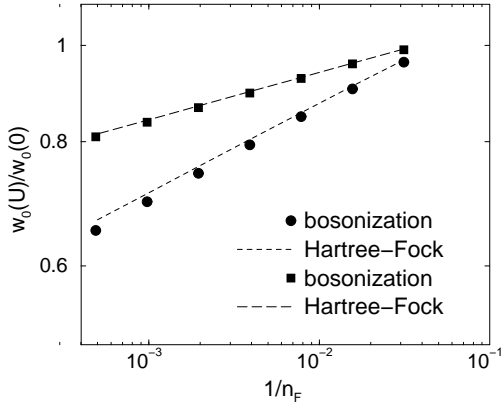}
\caption{Spectral weight at $\mu$
  and close to the boundary. 
  The circles show the exact results 
  for the spinless TL model for $x=3R$,
  $U/(2 \pi v_F) =0.1$ and $k_F = 4 q_c$. The long dashed 
  line presents the HF
  approximation. The squares show the exact results 
  for $x=3R$ and $U/(2 \pi v_F) =0.05$ and the dashed line the
  corresponding HF approximation.}
\label{fig4}
\end{center}
\end{figure}

\noindent For repulsive interactions ($\tilde{V}(0)>0$) the prefactor of the
logarithm is positive and the perturbative expression indicates a
suppression of the weight.   
To further investigate the behavior of $\rho(x,\omega)$ close to the
chemical potential 
we have studied the spectral weight at the chemical potential and 
position $x$, denoted    
$w_0(x,n_F;\tilde{V})$, for a given $k_F$ as a function of $1/n_F$
($\propto 1/L$).
The ratio $w_0(x,n_F;\tilde{V})/ w_0(x,n_F;0)$ displays the same 
kind of power-law behavior as does 
$\rho(x,\omega)$ as a function of $\omega$. 
Smaller systems are sufficient to find power-law behavior in
\linebreak 
$w_0(x,n_F;\tilde{V})/ w_0(x,n_F;0)$  compared to $\rho(x,\omega)$.
Fig.\ \ref{fig4} shows a log-log plot of 
$w_0(x,n_F;\tilde{V})/ w_0(x,n_F;0)$ 
for the exact solution and the HF approximation.
A power-law fit of the symbols gives the expected exponent Eqs.\
(\ref{alphaBdef}) and (\ref{LLparameterTL}) with high accuracy. 
Surprisingly also the HF approximation displays power-law behavior, 
which shows
that the leading logarithmic divergence of $\rho^{\rm HF}$ found in 
Eq.\ (\ref{specpertTL})
can be resummed to give a power-law. A detailed
study shows that $\alpha_B^{\rm HF}=\tilde{V}(0)/(2 \pi v_{\rm HF})$ 
and thus $\alpha_B^{\rm HF}$ and 
$\alpha_B$ do agree up to leading order in   
$\tilde{V}(0)/( 2 \pi v_F) $.
Quantitative agreement between exact results and HF can be reached for 
$ \tilde{V}(0) / (2 \pi v_F) \ll 1$.

\section{Lattice model of spinless fermions}
\label{spinless}

Next we consider the lattice model of spinless fermions
with $N$ lattice sites, lattice constant $a=1$, 
hopping matrix element $t = 1$, 
nearest neighbor interaction $U$, and OBC
\begin{eqnarray}
\hat{H}  = - \sum_{j=1}^{N-1} \left( c_j^{\dag} c_{j+1}^{} +
  c_{j+1}^{\dag} c_j^{}  \right)
  + U \sum_{j=1}^{N-1} n_j n_{j+1} .
\label{spinlessfermdef}
\end{eqnarray}
$c_{j}^{(\dag)}$ denotes the creation (annihilation) operator at 
site $j$ and $n_j =c_{j}^{\dag} c_{j}^{} $.
For $U=0$ the eigenstates of $\hat{H}$  are given by Eq.\
(\ref{eigenstates}) with $L \to N+1$, $x \to j$ and
$k_n = n\pi/(N+1)$,  $n \in \{1,2, \ldots, N \}$. Similar to PBC the
one-particle dispersion is $\varepsilon(k) = -2 \cos{(k)}$.
In contrast to the TL model the
interaction is of short range (in real space) and it is not obvious
whether the bosonization result Eq.\ (\ref{alphaBdef})  
$\alpha_B=K_{\rho}^{-1}-1$ holds. 
As for PBC the interacting model with OBC can be solved exactly by 
the Bethe ansatz, but 
similar to PBC not much about correlation functions 
can be learned directly from the solution. 
Information about boundary exponents can be obtained 
if conformal invariance is assumed\cite{WVP,BBHN}.

\begin{figure}[thb]
\begin{center}
\epsfxsize6.7cm
\epsfbox{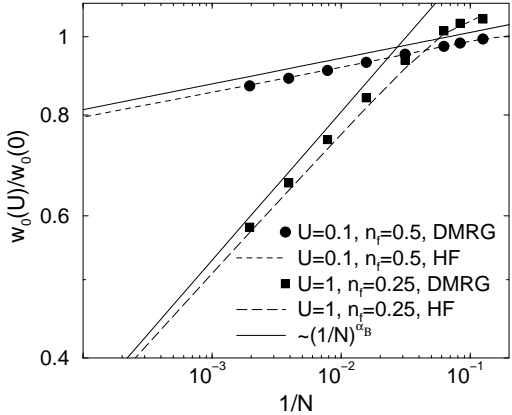}
\caption{Spectral weight at $\mu$
  and lattice site 1 for the lattice model of spinless fermions. 
  The symbols show the DMRG data and the
  dashed lines the corresponding HF results. For comparison the solid
  lines are power-laws with exponent $\alpha_B=K_{\rho}^{-1} -1$.
  The parameters are given in the legend.}
\label{fig5}
\end{center}
\end{figure}

Here we discuss the local spectral function at site $j=1$. We have
performed a DMRG study\cite{dmrg} for chains of up to $N=512$ sites 
calculating matrix elements
$w_0(n_F;U)=|\langle E^{n_F-1}_0|c_1|E^{n_F}_0 \rangle|^2$, i.\ e.\ the 
spectral weight at the chemical potential
and the boundary site (see Sec.\ \ref{TLM}). $|E^{n_F}_0 \rangle$
denotes the exact $n_F$-particle ground state. 
Results for $w_0(n_F;U)/w_0(n_F;0)$ 
as a function of $1/N$ (instead of \linebreak $1/n_F$), two different filling 
factors $n_f=n_F/N$, and $U$ are shown as the symbols in Fig.\
\ref{fig5}. $K_{\rho}(U,n_f)$ for these parameters can e.g.\ be found in 
Ref.\ \cite{Qin}: $K_{\rho}(U=0.1,n_f=0.5)=0.9691$ and  
$K_{\rho}(U=1,n_f=0.25)=0.8447$.
For large $N$ the numerical data nicely follow the solid lines, which are
proportional to power-laws with exponent $\alpha_B(U=1,n_f=0.25)=0.1838$ 
respectively $\alpha_B(U=0.1,n_f=0.5)=0.0319$.
The numerical error of the DMRG data is smaller than the size of the
symbols. 
We can thus conclude that the spectral weight close to the boundary and
the chemical potential shows a suppression with a power-law and 
exponent $\alpha_B$. This is consistent with
the prediction of bosonization and  results 
presented in Ref.\ \cite{WVP} 
obtained from Bethe ansatz and boundary conformal field theory.

In contrast to the TL model the HF problem for spinless fermions 
is numerically best studied in the site representation. The HF or mean-field
Hamiltonian is given by
\begin{eqnarray}
\hat{H}^{\rm HF} & = & - \sum_{j=1}^{N-1} \left(t(j) c_j^{\dag} c_{j+1}^{} +
 t^{\ast}(j) c_{j+1}^{\dag} c_j^{}  \right) \nonumber \\
 &&  + \sum_{j=1}^{N-1} [U+u(j)] n_j 
\label{spinlessfermHF}
\end{eqnarray}
with a renormalized hopping
\begin{eqnarray}
t(j) & = & 1+ U \langle c_{j+1}^{\dag} c_j  \rangle_0 \nonumber \\
& = & 1 +\frac{2 U}{N+1} \sum_{n=1}^{n_F} \sin{(k_n j)} \sin{(k_n
  [j+1])} ,
\label{renhopdef}
\end{eqnarray}
onsite energies 
\begin{eqnarray*}
u(j) = U   
\left\{  \begin{array}{cl}
           \langle n_2  \rangle_0   & \mbox{for} \,\, 
            j = 1  \\
            \langle n_{j-1}  \rangle_0
          +\langle n_{j+1}  \rangle_0
          & \mbox{for} \,\, j \in \{2,3, \ldots, N-1 \} \\
        \langle n_{N-1} \rangle_0   & \mbox{for} \,\, 
            j = N ,
\end{array} \right. 
\end{eqnarray*}
and
\begin{eqnarray}
\langle n_j  \rangle_0 = \frac{2}{N+1} 
 \sum_{n=1}^{n_F} \sin^2{(k_n j)} ,
\label{expecdef}
\end{eqnarray}
where $\langle \ldots \rangle_0$ denotes the
noninteracting ground state expectation value.
Due to the OBC the site occupation 
$\langle n_j \rangle_0 $ depends on $j$ showing Friedel oscillations. 
The sums in Eqs.\
(\ref{renhopdef}) and (\ref{expecdef}) can be performed analytically
and in order to obtain the site diagonal Green's 
function \linebreak 
$\left[G^{\rm HF}(\omega)\right]_{j,j}$, and thus the spectral 
function $\rho_j^{\rm HF}(\omega)$,
one numerically has to determine the eigenvalues and
eigenvectors of a tridiagonal 
matrix. This can easily be done for systems of up to 
$N=10^6$ lattice sites. 
Fig.\ \ref{fig5} shows HF data for $w_0(n_F;U)/w_0(n_F;0)$
and the same parameters as above. Similar to the TL model 
already the HF approximation displays power-law behavior with exponent 
$\alpha_B^{\rm HF}=\tilde{V}_{\rm eff}/
(2 \pi v_{\rm HF})$ which has the same form as for the TL model 
if one replaces $\tilde{V}(0)$ by the effective interaction 
$\tilde{V}_{\rm eff} = \tilde{V}(0)-\tilde{V}(2k_F) = 
2 U [1-\cos{(2k_F)}]$. This replacement is familiar from the mapping
to the ``$g$-ology'' model for PBC\cite{Solyom}: For spinless 
fermions forward $g_2$ and backward $g_1$ scattering are 
indistinguishable which leads to an effective coupling $g_2 - g_1$.
Again $\alpha_B$ and $\alpha_B^{\rm HF}$ agree up to order 
$\tilde{V}_{\rm eff}/(2 \pi v_F) $ and  
both curves show quantitative agreement for 
$\tilde{V}_{\rm eff}/(2 \pi v_F)  \ll 1$.

\section{Hubbard model}
\label{hubbard}

The main part of our earlier publication\cite{earlier} 
on LL's with boundaries is
devoted to a numerical investigation of the spectral properties of
the 1D repulsive Hubbard model
\begin{eqnarray}
\hat{H} &  = & - \sum_{s} \sum_{j=1}^{N-1} 
  \left( c_{j,s}^{\dag} c_{j+1,s}^{} + c_{j+1,s}^{\dag} c_{j,s}^{}  \right)
\nonumber \\  
&& + U \sum_{j=1}^{N-1} n_{j,\uparrow} n_{j,\downarrow} .
\label{hubbarddef}
\end{eqnarray}
Similar to the lattice model of spinless fermions 
Eq.\ (\ref{spinlessfermdef}) we set the lattice
constant $a$ and the hopping matrix element $t$ equal to unity. 
Here we will give a short summary of the numerical 
results obtained in Ref.\ \cite{earlier}, discuss new 
data sets, and then present an
analytical investigation of the crossover behavior of the spectral weight
within an effective low-energy model. 

\subsection{Numerical results: DMRG}
\label{hubbardDMRG}

\begin{figure}[thb]
\begin{center}
\epsfxsize6.7cm
\epsfbox{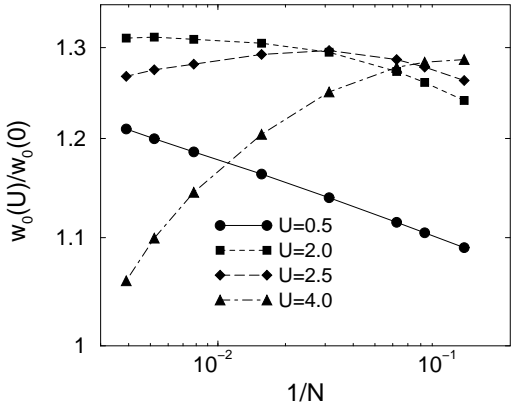}
\caption{DMRG results for the spectral weight at $\mu$
  and lattice site 1 for the Hubbard model at quarter filling. 
  The interaction $U$ is given in the legend.}
\label{fig6}
\end{center}
\vspace{-0.0cm}
\end{figure}

\begin{figure}[thb]
\begin{center}
\epsfxsize6.7cm
\epsfbox{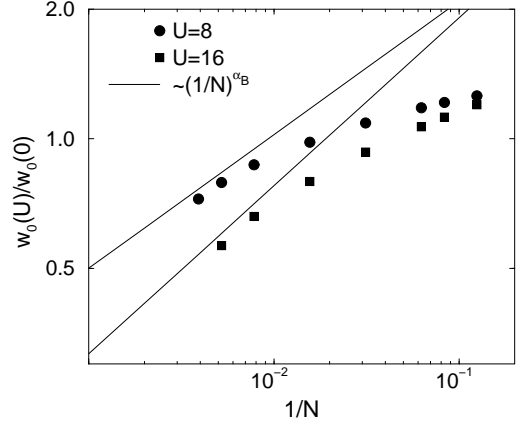}
\caption{The same as in Fig.\ \ref{fig6}, but for larger $U$. 
The solid lines are power-laws with exponent 
$\alpha_B=(K_{\rho}^{-1} -1)/2$.}
\label{fig7}
\end{center}
\end{figure}

As for the lattice model of spinless fermions we have calculated the
spectral weight at the boundary site and the chemical potential as a
function of the number of lattice sites $N$ using the DMRG algorithm.
We were able to obtain results for up to $N=256$ sites. Figs.\
\ref{fig6} and \ref{fig7} show data for
$w_0(n_F;U)/ w_0(n_F;0)$, quarter filling $n_f=2 n_F/N=0.5$,
and different $U$. The numerical error of the DMRG data is of the
order of the symbol size.
Instead of decreasing, as predicted by
bosonization and in contrast to our findings for the other two models
considered, 
$w_0(n_F;U)/ w_0(n_F;0)$ {\it increases} for small and moderate values of $U$ 
(see Fig.\ \ref{fig6}). For moderate $U \approx 2$ a crossover to 
a suppression occurs at system sizes 
reachable within DMRG. For smaller $U$, e.g.\ $U=0.5$, only the increase 
can be seen. We expect that the crossover sets in at much larger chain 
length.  
Only the large $U$ data of Fig.\ \ref{fig7} display a clear
suppression of the spectral weight for all the system sizes available
and thus only for these data a comparison to the power-law 
with exponent $\alpha_B=(K_{\rho}^{-1} -1)/2$ predicted by
bosonization seems meaningful.  
$K_{\rho}(U,n_f)$ for the Hubbard model has been calculated in 
Ref.\ \cite{Schulz}. For the parameters chosen we have  
$K_{\rho}(U=8,n_f=0.5)=0.62$ and  
$K_{\rho}(U=16,n_f=0.5)=0.56$. Power-laws with exponents 
$\alpha_B(U=8,n_f=0.5)=0.31$ and $\alpha_B(U=16,n_f=0.5)=0.39$
are shown as solid lines in Fig.\ \ref{fig7}.
For large $N$ the numerical data seem to approach these lines and we 
conclude that the DMRG results are consistent with a
final power-law suppression of the spectral weight near the boundary 
which is given by the boundary exponent  $\alpha_B $. 
The asymptotic behavior is thus consistent with the prediction of 
bosonization\cite{FG,EMJ} and results obtained using Bethe ansatz and
boundary conformal field theory\cite{WVP}.  

The surprising new finding is the increase of weight 
for small and moderate $U$
and the
subsequent crossover. 
This could have been expected already from the lowest order 
result Eq.\ (\ref{specpert}) for the spectral function.
For a $k$-independent interaction $U$  
the leading logarithmic correction to $\rho^{H}$ is given by\cite{HnotHF}
\begin{eqnarray}
\rho^{H}(x,\omega)  \sim   1 + 
\left(1-z \right)  \frac{U}{2 \pi v_F} \, \ln  
   \left|\frac{\omega}{v_F k_F} \right| +  \ldots \,\, . 
\label{specpertH}
\end{eqnarray}  
The prefactor of the logarithmic correction 
in a model with spin ($z=2$) has thus the opposite sign as in the case of a 
long range interaction Eq.\ (\ref{specpertTL}).
As long as $[U/ (2 \pi v_F)] \ln{(N)}  \ll 1$
this indicates a logarithmic increase of the weight displayed in
Fig.\ \ref{fig6}. 
The crossover scale can be estimated from this
expression to be $1/N^A_c \sim \exp{[- 2 \pi v_F / U]}$.
From the data it is clear that also the exact 
crossover scale $1/N_c$ strongly depends on $U$. Going from 
$U=2.5$ to $U=2$ it roughly decreases by one order of magnitude.
As a consequence of this and the fact that
we are limited to $N=256$ lattice sites the crossover can only be
observed within a small window of interactions $2 \leq U \leq 3$.
Below we will show that the crossover scale indeed 
decreases exponentially in $-1/U$ using our analytical Hartree result 
and scaling arguments. The $1/N$ dependence of
$w_0(n_F;U)/ w_0(n_F;0)$ can be translated into the $\omega$
dependence of $\rho_1(\omega)$ by multiplying by the inverse
of the noninteracting density of states $\pi
v_F$. For the energy scale at which the decrease of the weight sets in
we thus find $\Delta = v_F \pi/N_c$. For e.g.\ the $U=2.5$ data at
quarter filling this leads to $\Delta/B \approx 10^{-2}$, where $B$
denotes the bandwidth $B=4$. This implies that for the given
parameters the prediction of bosonization only holds for
energies much smaller than one hundredth of the bandwidth.
In Sec.\ \ref{summary} we will further discuss this issue.

\subsection{Numerical results: Hartree approximation}
\label{hubbardnumHartree}

Numerically the Hartree approximation (the Fock term vanishes) for the
Hubbard model is best studied in the site representation. 
It can be done in close analogy to the 
lattice model of spinless fermions discussed in Eqs.\
(\ref{spinlessfermHF}) to (\ref{expecdef}).
For the Hubbard model the hopping is not modified by the interaction 
and the 
one-particle problem
which remains to be solved for an electron of spin species $s$  
is the dynamics in an external  
Hartree potential generated by $\langle n_{j,-s} \rangle_0$. The
Hartree potential  shows Friedel oscillations, i.e.\ 
oscillates as $\cos{(2 k_F j)}$ and slowly decays as $1/j$.  
In the continuum limit external potentials of this form are called
Wigner-von Neumann type potentials\cite{mathpaper}. 
From scattering theory it is known that the eigenvalues and
eigenstates of Hamilton operators with  oscillating and slowly
decaying ($\sim 1/x$) external potentials have
unusual properties\cite{mathpaper}. Numerically the spectral weight of
the lattice model can be calculated along the lines discussed in the
last section. 
In Figs.\ 2 and 3 of  Ref.\ \cite{earlier} we have presented
numerical data for $\rho_1^{ H}(\omega)$ as a function of $\omega$, 
different $U$, and system sizes of up to $N=10^6$ lattice sites. 
We have restricted ourselves to energies close to the chemical potential.
For small and moderate $U$ the spectra
display qualitatively the same behavior as the DMRG data discussed
above. For energies 
approaching the chemical potential from below 
the weight first increases following a power-law.
From the numerical data the exponent can be determined with high
accuracy to be $-U/(2 \pi v_F)$.
This shows that the logarithmic divergence Eq.\
(\ref{specpertH}) obtained in the leading order solution of the Hartree
problem can be resummed to produce a power-law in an energy range
close to the chemical potential. The numerical Hartree data then
display a crossover on a scale which decreases exponentially in $-1/U$,
and a subsequent power-law suppression. From the numerical data the exponent 
is found to be
\begin{eqnarray}
\alpha_B^H = U/(2 \pi v_F) .
\label{alphaBHdef}
\end{eqnarray}
$\alpha_B^H$ has to be compared to the
leading behavior of the exact boundary exponent 
$\alpha_B = (K_{\rho}^{-1} -1)/2$. It is given by 
$\alpha_B = U/(4 \pi v_F) 
+ {\mathcal O}\left( [U/2 \pi v_F]^2\right)$\cite{missprint}  
which is one-half of $\alpha_B^H$. 
This kind of discrepancy between exponents obtained 
in perturbation theory and 
the leading behavior of exact exponents is known from the 
Hubbard model with PBC\cite{Schliecker}. It 
occurs because the scaling of coupling constants (of the backscattering
contribution in case of PBC) is not taken into
account in simple perturbation theory. Thus the coupling constant
relevant for exponents is overestimated in a naive perturbative
treatment. For the same reason DMRG and Hartree data only show 
qualitatively agreement even in the limit of small coupling 
(see Fig.\ 4 of Ref.\ \cite{earlier}). This has to be
contrasted to the above results for the TL model and the lattice model
of spinless fermions. 
In analogy to PBC the couplings in these two
models do not scale\cite{Solyom}. 

\begin{figure}[thb]
\begin{center}
\epsfxsize6.7cm
\epsfbox{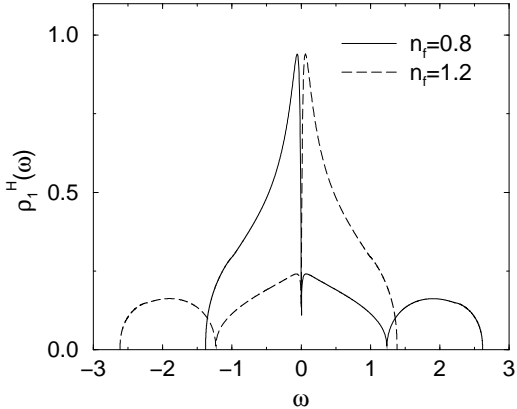}
\caption{Spectral function of the Hubbard model 
at site $j=1$, for $N=2000$, $U=5$, and two different filling factors
$n_f$ in the Hartree approximation.}
\label{fig8}
\end{center}
\end{figure}

Fig.\ \ref{fig8} shows $\rho_1^{H}(\omega)$ as a
function of $\omega$ for all energies within the band, $N=2000$, 
$U=5$, and two filling factors $n_f$ which are related by $n_f \to
2-n_f$. The individual weights have been connected to a continuous
line.  Since the hopping amplitude is not 
renormalized by the interaction the total bandwidth 
is equal to the noninteracting one $B=4$.  
Besides the crossover behavior for energies
close to the chemical potential, i.e.\ $\omega=0$,
the solid line with $n_f=0.8$ shows a symmetric suppression of 
weight around $\omega = - 2 \varepsilon(k_F) >0 $, i.e.\ at energies 
which are unoccupied in the ground state. On the Hartree level the
spectral function shows two important symmetries. 
In the next subsection the reason for both, the supression at 
$\omega = - 2 \varepsilon(k_F)$ and the symmetries will become clear.
The spectral 
function for fixed filling and repulsive interaction $U>0$ can be 
mapped onto the one for attractive interaction with the same 
absolut value $-|U|$ and the same filling by taking the mirror 
image at $\omega = - \varepsilon(k_F)$. In this way the weight 
for $U>0$ around $\omega = - 2 \varepsilon(k_F)$ is mapped onto the one for 
$-|U|$ around $\omega=0$, i.e.\ the chemical potential. 
Here we do not consider the spectral function for 
negative $U$ any further since the Hubbard model with attractive 
interaction is not a LL\cite{Solyom,Voit}. 
The second symmetry is obvious from Fig.\ \ref{fig8}. The spectral
function for fixed $U$ and filling $n_f <1$ can be mapped onto the one
with filling $2-n_f$ by taking the mirror image at $\omega=0$. 
Note that the weight for $U>0$ and energies around the chemical 
potential shows a strong asymmetry. It implies that for $n_f<1$ 
the increase preceding the final suppression of the weight
of energies which are occupied in
the ground state is more pronounced compared to the analogous weight for 
fillings $n_f>1$. This has to be contrasted to 
bosonization which always gives symmetric behavior around 
$\omega=0$.  

Next we will present an analytical
discussion of the spectral weight within the Hartree approximation. 
We will be able to analytically determine the exponents of the 
power-law increase, the subsequent decrease and the cross\-over scale
$\Delta(U,n_f)$ discussed above.

\subsection{Analytical results}
\label{hubbardanaHartree}

\begin{figure}[thb]
\begin{center}
\epsfxsize5.0cm
\epsfbox{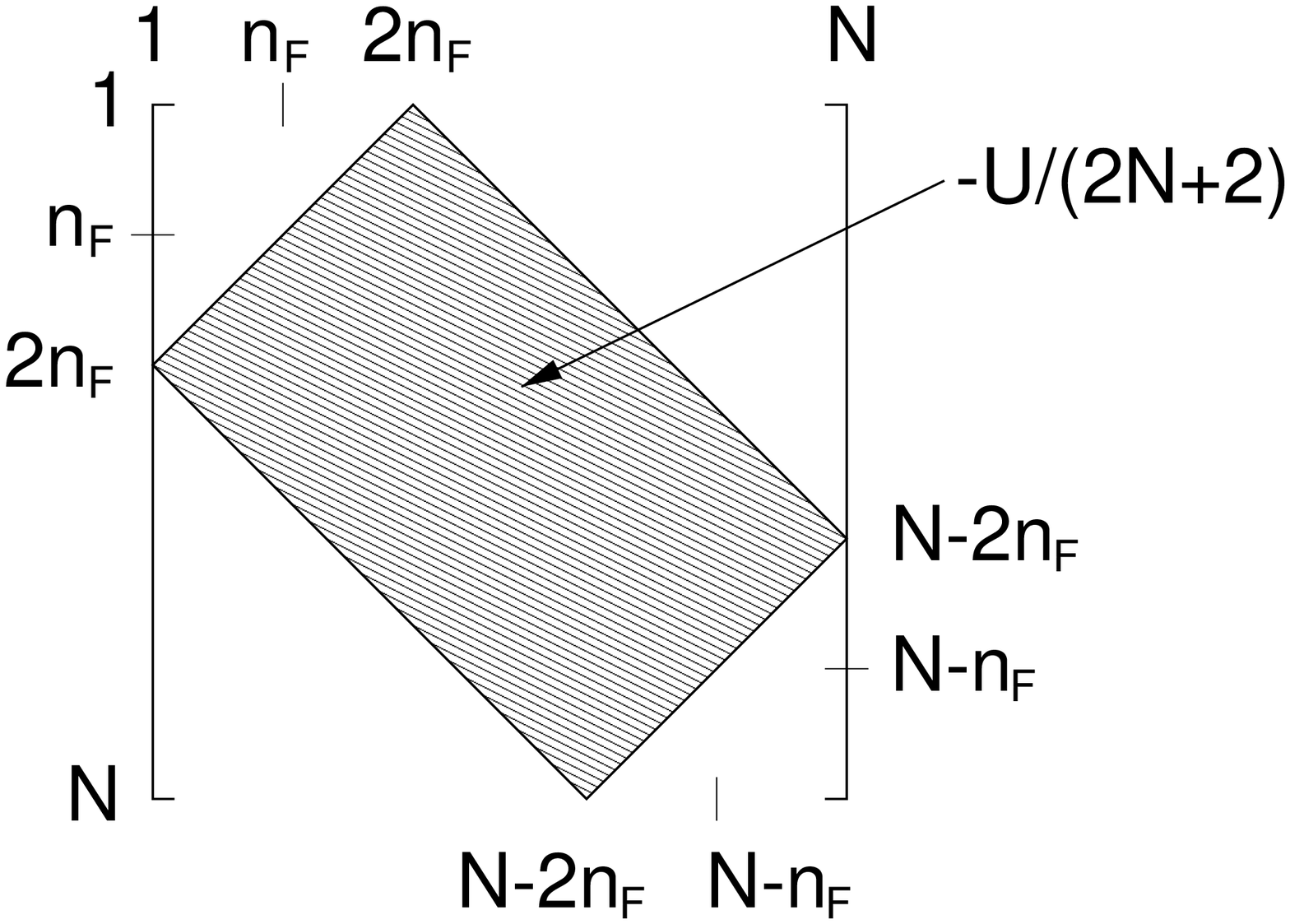}
\caption{The self-energy matrix $\tilde{\Sigma}^{H}_s$ for the 
Hubbard model with $n_f<1$. For details see the text.}
\label{fig9}
\end{center}
\end{figure}

For an analytical discussion of the Hartree approximation 
it is preferable to work in
$k$-space. Similar to the case of a general continuum model 
Eqs.\ (\ref{interactionII}) to (\ref{Fcalculated}) we 
thus first calculate the matrix elements of the local Hubbard 
interaction in the noninteracting eigenstates
\begin{eqnarray}
\varphi_n (j) = \sqrt{\frac{2}{N+1}} \sin{ (k_n j)} ,
\label{Heigenstates}
\end{eqnarray}
with $k_n = n\pi/(N+1)$,  $n \in \{1,2, \ldots , N \}$. 
They can be written similar to Eq.\ (\ref{matrixelementII}) with 
\begin{eqnarray}  
F_{s,s'}(q_m,q_{m'}) & = & \delta_{s, -s'} 
\frac{U}{N+1} \sum_{j=0}^{N} \cos{(q_m j)} 
\cos{(q_{m'} j)}  \nonumber \\
& = & \delta_{s, -s'}  \frac{U}{2} \left( \delta_{m,m'}^{(N+1)} +
  \delta_{m,-m'}^{(N+1)}  \right) , 
\label{FdefHubbard}
\end{eqnarray}
and the $(N+1)$-periodic Kronecker $\delta$, $\delta_{m,m'}^{(N+1)} =1 $ for 
$m = m' + l (N+1)$, $l \in {\bf Z}$ and zero otherwise. 
A comparison with Eq.\
(\ref{Fcalculated}) shows that the correction term $g(q,q')$
vanishes for a local interaction. From Eq.\  (\ref{FdefHubbard}) the
Hartree self-energy can be calculated. 
In Fig.\ \ref{fig9} the self-energy matrix   
\begin{eqnarray}
\tilde{\Sigma}^{H}_s = \Sigma^{H}_s 
- \delta \mu  {\mathbf 1} 
\label{Stildedef}
\end{eqnarray}
with $\delta \mu = U (n_F +1/2)/(N+1)$, 
is sketched for a filling factor $n_f< 1$. 
Similar to the TL model the function $f([n+n']/2)$  Eq.\
(\ref{kleinfdef}) enters the expression for the self-energy and 
only matrix elements with even $n+n'$ do contribute.
In the hatched area and for even $n+n'$ the 
matrix elements are given by $-U/[2(N+1)]$. They vanish outside this
area. 
For half filling $n_f=1$ all the off-diagonal
elements of the self-energy matrix 
vanish and the spectral function is given by the
noninteracting one. For fillings $n_f>1$,
$\tilde{\Sigma}^{H}_s$ 
is given as sketched in Fig.\ \ref{fig9} 
but with $n_F \to N-n_F$ and $U \to -U$. This leads to the symmetries 
of the spectral weight discussed in the last subsection. 
From now on we will restrict ourselves to fillings
$n_f<1$. The weight for $n_f>1$ can be constructed using the discussed
symmetry.  
From the self-energy the Green's function 
$\left[G(\omega)\right]_{n,n'}$ can be calculated
following Eq.\ (\ref{Gretard}).

A lowest order inversion of $\left[ G^{H} \right]^{-1}$ leads to Eq.\
(\ref{specpertH}). The calculation shows that the nonanalytic
behavior of the spectral function close to the chemical potential,
i.e.\ close to $\omega=0$, 
comes from the step through the point $(n_F,n_F)$ 
in the upper left part of the self-energy matrix 
Fig.\ \ref{fig9}. It is given by the line $n'=2n_F+1-n$. 
The sharp step in the lower right half, given by 
$n'= 2(N+1)-2n_F -1 -n$, comes from  umklapp processes 
where $m=m'+N+1$ in Eq.\ (\ref{FdefHubbard}). 
In the spectral function it leads to the symmetric 
suppression of the weight close to the unoccupied energy 
$\omega = -2 \varepsilon(k_F)$ 
already discussed in the last subsection 
(see the solid line in Fig.\ \ref{fig8}).
Numerically it can be shown that the 
suppression is given by a power-law with exponent $U/(2 \pi v_F)$.


\subsubsection{Effective model}

As we are only interested in the nonanalytic behavior of the spectral
weight for energies close to the chemical potential the Hartree
problem can be simplified. This leads to an effective low-energy
model. Due to the structure of the self-energy matrix the  
Hartree problem separates into two equivalent problems. 
One for even $n$ and $n'$ and the other one for odd 
$n$ and $n'$. In both problems the vanishing matrix elements in
the hatched area of Fig.\ \ref{fig9} are left out, but the level
spacing is doubled. Equivalently one can work with a ``coarse grained'' 
version of $\tilde{\Sigma}^{H}_s$. It has nonvanishing matrix elements 
for all $n$ and $n'$ in the hatched area of Fig.\ \ref{fig9}, but 
they have only half the size of the original ones. 
As only the sharp step in the upper left part is
important for the nonanalytic behavior close to $\mu$ 
the matrix $\tilde{\Sigma}^{H}_s$ can be replaced by a matrix
$\mathcal V$ with elements
\begin{eqnarray}
{\mathcal V}_{n-n_F,n'-n_F} 
= -\frac{U}{4(N+1)} \, \Theta(n+{n'}-2n_F) . 
\label{Vdef}
\end{eqnarray}
Next the space of one-particle states is changed. 
Without modifying the low-energy properties it can be 
reduced to $n \in \{1,2, \ldots 2 n_F \}$. 
Furthermore the one-particle dispersion $\xi(k)$ can be linearized 
around $k_F$
\begin{eqnarray}
\xi(k-k_F) = v_F (k-k_F) .   
\label{Xilindef}
\end{eqnarray}
In a last step the quantum numbers $n$ are shifted by $n_F$.
We then have to determine the local spectral function for a scattering
problem with the Hamiltonian
\begin{eqnarray}
\hat{H}_{\rm eff}  = \sum_{n=-n_F+1}^{n_F} v_F k_n 
a^\dagger_{n} a^{}_{n} + \sum_{n,n'=-n_F+1}^{n_F} 
{\mathcal V}_{n,n'} \, a^\dagger_{n} a^{}_{n'} .
\label{Heffdef}
\end{eqnarray}
In order to obtain a continuous spectral function we take the
thermodynamic limit. This is accomplished by switching to unperturbed
one-particle states 
\begin{eqnarray}
\tilde{\varphi}_k (j) = \sqrt{\frac{2}{\pi}} \sin{ [(k_n+k_F) j]} .
\label{phihere}
\end{eqnarray}
and using 
\begin{eqnarray}
\sum_{n} \longrightarrow \frac{N+1}{\pi} \int d \, k .
\label{thlimit}
\end{eqnarray}
In these states the scattering potential is given by 
\begin{eqnarray}
{\mathcal V}_{k,k'} 
= -\frac{U}{4 \pi} \, \Theta(k+k') . 
\label{Vkstatesdef}
\end{eqnarray}


\subsubsection{Scattering theory}

As usual in scattering theory\cite{Taylor} 
we define the off-shell $T$ matrix $T(\omega)$
which obeys the Lippmann-Schwinger equation
\begin{eqnarray}
T(\omega) = {\mathcal V} + {\mathcal V} G_0(\omega) T(\omega) ,
\label{LippSchwingT}
\end{eqnarray}
with the bare retarded Green's function 
\begin{eqnarray}
\left[G_0(\omega)\right]_{k,k'} = \frac{1}{\omega -
  v_F k +i0} \, \delta(k-k') .
\label{barepropdef}
\end{eqnarray}
The local spectral function can then be calculated from
Eq.\ (\ref{specdef}) and 
\begin{eqnarray}
G(\omega) = G_0(\omega) +  G_0(\omega) T(\omega) G_0(\omega) .
\label{GfromT}
\end{eqnarray} 
For the present problem it turns out to be advantageous to
express the Green's function in terms of the so called $K$ (or
Heitler) matrix\cite{Taylor}
which obeys the Lippmann-Schwinger equation
\begin{eqnarray}
K(\omega) = {\mathcal V} + {\mathcal V} G_0^{R}(\omega) K(\omega) ,
\label{LippSchwingK}
\end{eqnarray}
with 
\begin{eqnarray}
\left[G_0^R(\omega)\right]_{k,k'} =  
\frac{\mathcal P}{\omega -  v_F k} \, \delta(k-k') .
\label{barepropRdef}
\end{eqnarray}
Here $\mathcal P$ denotes the principal value. $T$ and $K$ are 
related by 
\begin{eqnarray}
\left[T(\omega)\right]_{k,k'} = \left[K(\omega)\right]_{k,k'} 
- i \pi 
\frac{\left[K(\omega)\right]_{k,\tilde{\omega}} 
\left[K(\omega)\right]_{\tilde{\omega},k'}}{v_F + i \pi 
\left[K(\omega)\right]_{\tilde{\omega},\tilde{\omega}}} ,
\label{TfromK}
\end{eqnarray}
with $\tilde{\omega} = \omega/v_F$.
Using Eqs.\ (\ref{GfromT}) and (\ref{specdef}) this leads to
\begin{eqnarray} 
\rho^{\rm eff}_j(\omega) =  \rho^{0}_j(\omega) \frac{h_j(\omega)}{1+
  \pi^2 g(\omega)} ,
\label{finaldef}
\end{eqnarray}
where the noninteracting spectral weight is given by Eq.\
(\ref{specnoint})
\begin{eqnarray} 
\rho^{0}_j(\omega) =  \frac{1-\cos{\left(2 
\left[ k_F + \omega/v_F \right] j \right) }}{\pi v_F}  .
\label{rho0def}
\end{eqnarray}
Close to the boundary and for $|\omega| \ll v_F k_F$,
$\rho^{0}_j(\omega)$ can be replaced by its value at $\omega=0$.
The functions $g$ and $h_j$ are given by 
\begin{eqnarray}
g(\omega) = \left[K(\omega)\right]^2_{\tilde{\omega},\tilde{\omega}}/v_F^2 
\label{gdef}
\end{eqnarray}
and 
\begin{eqnarray}
h_j(\omega) = \left[ 1+ {\mathcal P} \! \! \! \! \! \! 
\int_{-k_F}^{k_F} d \, k 
\frac{\tilde{\varphi}_{k}(j)}{\tilde{\varphi}_{\tilde{\omega}}(j)} 
\frac{\left[K(\omega)\right]_{k,\tilde{\omega}}}{\omega - v_F k} 
\right]^2 , 
\label{hdef}
\end{eqnarray}
with
$\tilde{\varphi}_{k}(j)$ as in Eq.\ (\ref{phihere}). 
The leading contribution to the integral in Eq.\ (\ref{hdef}) comes
from the pole at $\omega = v_F k$.  
For small $j$, i.e.\ lattice sites close to the boundary, 
small $k$ and small $\omega$, 
$\tilde{\varphi}_{k}(j)/\tilde{\varphi}_{\tilde{\omega}}(j)$ depends only  
weakly on $k$ and can thus be replaced by unity. 


\subsubsection{Spectral function}

In the Appendix we partly solve the Lippmann-Schwinger equation 
for the off-shell $K$ matrix Eq.\ (\ref{LippSchwingK}) and present
arguments which strongly suggest  
that the leading behavior of $g(\omega)$ 
and $h_1(\omega)$ for small $|\omega|$ is given by power-laws
\begin{eqnarray}
g(\omega) = \left[\frac{U}{8 \pi v_F} \right]^2 \left[ 
\left| \frac{2 \omega}{v_F k_F} \right|^{- \alpha_B^H}  + \mbox{sign} 
(\omega) \right]^2 ,
\label{geval}
\end{eqnarray}
and
\begin{eqnarray}
h_1(\omega) = 
\left| \frac{2 \omega}{v_F k_F} \right|^{- \alpha_B^H} .
\label{heval}
\end{eqnarray}
The boundary exponent $\alpha_B^H$ is defined in Eq.\ (\ref{alphaBHdef}). 
Using Eq.\ (\ref{finaldef}) 
for $|\omega| \ll v_F k_F$  the spectral function at the boundary site 
is thus given by  
\begin{eqnarray}
\rho^{\rm eff}_1(\omega) \sim 
\frac{\left| \frac{2 \omega}{v_F k_F} \right|^{- \alpha_B^H}}
  {1+  \left[\frac{U}{8 v_F} \right]^2 \left[ 
\left| \frac{2 \omega}{v_F k_F} \right|^{- \alpha_B^H}  + \mbox{sign} 
(\omega) \right]^2}.
\label{complete}
\end{eqnarray}
The way the effective model was
constructed this analytical result
captures the leading small $|\omega|$ behavior
of the local spectral function of the  
Hubbard model within the Hartree approximation. 
All the features found in our numerical calculations Sec.\ 
\ref{hubbardnumHartree} and Ref.\ \cite{earlier} are 
confirmed by Eq.\ (\ref{complete}).
For $U>0$ and $|\omega| \to 0$ we first find a power-law increase 
of the weight with exponent $-\alpha_B^H$ given by the numerator of
Eq.\ (\ref{complete}). For even smaller $|\omega|$ the denominator
becomes important and a power-law decrease with $\alpha_B^H$ sets in. From
Eq.\ (\ref{complete}) the crossover scale $\Delta$ 
can be calculated analytically 
\begin{eqnarray}
\frac{\Delta}{v_F k_F} = \exp{\left\{ -  \frac{\pi v_F}{U} 
\ln{\frac{1+ [U/(8 v_F)]^2}{ [U/(8 v_F)]^2}} \right\}} .
\label{crossoverscale}
\end{eqnarray}
Up to logarithmic corrections it is exponentially small in $-1/U$ as
already expected from the numerical data. The $\mbox{sign} 
(\omega)$ function in the denominator of Eq.\ (\ref{complete}) is responsible
for the asymmetry of the spectral weight around $\omega=0$ discussed in 
Sec.\ \ref{hubbardnumHartree}.

The analytical result Eq.\ (\ref{complete}) makes clear that it was
essential to determine the leading small $|\omega|$ behavior of the $K$
matrix Eq.\ (\ref{LippSchwingK}), respectively of $g(\omega)$ Eq.\
(\ref{gdef}) and $h_j(\omega)$ Eq.\ (\ref{hdef}), instead of the
leading behavior of the $T$ matrix: The latter one gives the final
power-law decrease, but does not capture the power-law increase and
the crossover. 

For $U<0$ only the numerator of Eq.\
(\ref{complete}) is relevant and leads to the symmetric power-law 
suppression of the weight close to $\mu$ with exponent 
$|U|/(2 \pi v_F)$.  This confirms the numerical results of 
Sec.\ \ref{hubbardnumHartree}.

We note in passing that by calculating the local spectral function 
of the 1D Hubbard model in the Hartree approximation, we have 
determined the local spectral function for the one-particle scattering 
problem of an electron in an external $\cos{(2  k_F x)}/x$ potential. 
The Jost function and scattering matrix for this problem have 
recently been discussed in the literature\cite{mathpaper}. 


\section{Discussion and Summary}
\label{summary}

Our study of the one-particle properties of LL's with 
open boundaries covers different aspects of the low-energy physics of
1D electrons with repulsive interaction. 
We have carefully reinvestigated the
bosonization approach to 1D models with OBC, especially focusing on
the subtleties, which go beyond the ones known from bosonization of
models with PBC. As no RG study exists which takes  all
the scattering processes into account we have restricted
ourselves to interactions which are long range in real space and the
low-energy subspace of the Fock space applying
Tomonagas original idea\cite{Tomonaga} to models with OBC. We were
then  able to confirm the result for the boundary exponent $\alpha_B$ 
which characterizes the suppression of spectral weight close to the
boundary and the chemical
potential\cite{FG}. $\alpha_B=(K_{\rho}^{-1}-1)/z$ can indeed be
expressed in terms of the bulk LL parameter $K_{\rho}$ of the
Tomonaga-Luttinger model.  Led by this observation and analogy to PBC 
several authors have generalized the above 
results to all models of LL's\cite{FG,EMJ}. Using the numerically exact
DMRG algorithm we investigated whether this is a legitimate
generalization for two models with a short range interaction: The
lattice model of spinless fermions with nearest neighbor interaction
and the 1D Hubbard model\cite{WVP,VWG}. In both cases we were able 
to explicitly verify that the final  
suppression of the spectral weight at small energies is
consistent with the prediction of bosonization, although for the
Hubbard model only at energies surprisingly close to the chemical
potential, respectively for very long chains.    

Besides that our study has revealed two interesting new   
results. Firstly  we found that many aspects of the influence 
a boundary has on the low-energy one-particle spectra can 
be understood within the Hartree-Fock approximation for the
self-energy. Using numerical and analytical techniques we were able 
to explain how the nontrivial structure of the self-energy matrix 
leads to power-law behavior of the Hartree-Fock spectral 
function close to the boundary. For the two models 
with dominant forward scattering (TL model and lattice model of
spinless fermions) quantitative agreement between the exact spectral
function and the approximated one can be reached for small
interactions. In both cases the exact boundary exponent 
$\alpha_B$ and $\alpha_B^{\rm
  HF}$ agree up to leading order in the interaction and the energy
range over which power-law behavior can be observed is large compared
to the one observed in the Hubbard model.  

The second surprising finding is the crossover behavior of the
spectral function of the Hubbard model. In both the numerically exact
DMRG calculation and the Hartree approximation the spectral weight
for small and moderate $U$ initially increases for energies 
approaching the chemical potential. Only below a crossover scale
$\Delta$ the power-law suppression predicted by bosonization sets in. 
The DMRG data reveal that $\Delta$ strongly decreases with
decreasing interaction $U$. Within the Hartree approximation we were 
able to show analytically that up to small corrections 
$\Delta = \Delta_0 \exp{(-U_0/ U)}$, with $U_0=\pi v_F$ and
$\Delta_0 = v_F k_F$. 
Due to the scaling of coupling
constants not captured in the Hartree approximation only qualitative
agreement between the DMRG and Hartree spectral weight can be
reached. The scaling is responsible for the fact that the Hartree
boundary exponent $\alpha_B^H$ is twice as large as the
leading behavior of the exact exponent. Furthermore the crossover scale 
is underestimated by the Hartree approximation (see Fig.\ 4 of Ref.\
\cite{earlier}). Nonetheless we believe that the exact crossover
scale shows an exponential $-1/U$ dependence, with a modified $U_0$ and
$\Delta_0$. Within a 
perturbative RG calculation similar to ``$g$-ology'' we expect that 
irrelevant couplings scale to zero as 
$1/ \left|\ln{\left( \Lambda/\Lambda_0 \right)} \right|$ if
$\Lambda \to 0$. 
Here $\Lambda$ denotes the momentum scale  
up to which degrees of freedom have been integrated out and
$\Lambda_0$ is the bare momentum cutoff (both measured relative to
$k_F$). Thus irrelevant couplings scale down on exponentially (in
$-1/U$) small scales. This implies that only $\Delta_0$ and $U_0$ 
of the exponential dependence of $\Delta$ found 
in the Hartree approximation are modified. A similar result can be
obtained if one starts at the LL fixed point and takes into account
the anomalous dimension of the fermion field. 
Close to the LL fixed point the leading irrelevant couplings are 
expected to scale to zero as $(\Lambda/\Lambda_0)^{\gamma(U)}$, with
$\gamma(U) \sim U$. This again leads to an exponential scale on which
irrelevant couplings scale down. Our observation has dramatic
implications for the energy range over which the prediction of 
bosonization is valid in the present context: It vanishes 
exponentially in $-1/U$. 

The perturbative result Eq.\ (\ref{specpert}) for the spectral weight 
indicates that the crossover behavior can be found in all models with
dominant total backscattering $z \tilde{V}(2 k_F) > \tilde{V}(0)$.

Over the past ten years several groups have attempted to 
find LL behavior in a variety of systems which behave as
quasi one-dimensional conductors using angular integrated
and angular resolved photoemission 
spectros\-co\-py\cite{Grioni,Zwick,Denlinger,Xue,Baer}. 
Here we do not want to give a complete account of all the experiments
and the subtleties of their interpretation\cite{Grioni} and focus on
one aspect, which we believe 
has to be reconsidered in the light of our results.  
The angular integrated spectra of all the possible candidates for 
LL behavior show a suppression of spectral weight close to the
chemical potential\cite{Grioni}. 
If the data are interpreted in terms
of the LL picture surprisingly large exponents $\alpha$ of the order
of 1 are required. Within bulk LL theory this implies  
$K_{\rho}$'s smaller then the once for which umklapp scattering  
becomes relevant taking the filling factors of the experimental 
systems. This leads to a contradiction since the data do not show signs
of a gap. 
Furthermore in a recent angular resolved 
measurement\cite{Zwick} the authors were
unable to detect any dispersion of the peaks measured. Motivated by
the theoretical expectation that a LL chain with impurities will,
at low energies, behave as if it is cut into several disconnected
pieces with open ends\cite{MattisII,KF} both the 
missing dispersion and the large exponent have been interpreted in 
the light of this expectation. 
If the electrons come from localized regions of
the chain, they do not display any dispersion.
For electrons coming from positions close to an open boundary
$\alpha_B$ would be the relevant exponent. As we have seen earlier
$\alpha_B > \alpha$. This would resolve the apparent contradiction
mentioned above, since a $\alpha_B$ of the order of 1 can still 
lead to a $K_{\rho}$ larger then the critical value at which umklapp
scattering leads to a gapped system. 
Thus it has been
speculated\cite{EMJ,VWG,Grioni,Zwick,Baer} that data have to be
compared with calculations for LL's with open ends. As we have
demonstrated above in these systems the energy range over which the LL
exponent $\alpha_B$ can be observed might be very small and taking
thermal and experimental broadening into account
even masked by a pronounced peak at
higher energies. At the lower end this energy range is furthermore cut
off by non LL effects as e.g.\ interchain hopping\cite{interchainhopping}.   
For a variety of reasons our results can certainly not directly be
applied to experimental spectra. One is that for the Hubbard
model studied boundary exponents $\alpha_B \approx 1$ cannot be
reached. What would thus be very desirable is a calculation of the
boundary spectral weight in a microscopic lattice model with an
interaction which is of longer range leading to boundary exponents of
the order of one.  We conclude that a convincing 
interpretation of the data in terms of boundary effects 
is still missing and requires further theoretical 
investigations. 


\begin{acknowledgement}
We would like to thank A.\ Rosch and J.\ Voit for
helpful discussions. This work has been supported by the SFB 341
of the Deutsche Forschungsgemeinschaft (V.\ M.\ and W.\ M.). 
Part of this work was performed while K.S.\ spent his sabbatical at
the ICTP in Trieste. He would like to thank  YU, Lu for the hospitality
in his group.
Part of the calculations 
were carried out on a T3E of the Forschungszentrum J\"ulich.
\end{acknowledgement}

\section*{Appendix}
\renewcommand{\theequation}{\mbox{A\arabic{equation}}}
\setcounter{equation}{0}

In this appendix we will determine the leading small $|\omega|$ 
behavior of the functions $g(\omega)$ and $h_1(\omega)$ defined in 
Eqs.\ (\ref{gdef}) and (\ref{hdef}) from the solution of 
the Lippmann- \linebreak Schwinger equation
(\ref{LippSchwingK}). 
For the dimensionless $K$ matrix $\tilde{K}(\omega) = K(\omega)/v_F$
it reads
\begin{eqnarray}
\left[\tilde{K}(\omega)\right]_{k,k'} = \tilde{\mathcal V}_{k,k'}+
 {\mathcal P} \! \! \! \! \! \!
\int_{-k_F}^{k_F} dp \frac{\tilde{\mathcal V}_{k,p}
  \left[\tilde{K}(\omega)\right]_{p,k'}}{\tilde{\omega} - p} , 
\label{LipSchwing}
\end{eqnarray}
with 
\begin{eqnarray}
\tilde{\mathcal V}_{k,k'}  = {\mathcal V}_{k,k'}/v_F 
=  - \tilde{U} \Theta(k+k') ,
\label{Vtildedef}
\end{eqnarray}
where $\tilde{U} = U/(4 \pi v_F)$, and $\tilde{\omega} = \omega/v_F$. 
By differentiating Eq.\
(\ref{LipSchwing}) with respect to $k$ a system of coupled 
differential equations for 
\begin{eqnarray} 
{\mathcal F}_1(k) \equiv {\mathcal F}_1(k;k',\omega) =  
\left[\tilde{K}(\omega)\right]_{k,k'} -
\tilde{\mathcal V}_{k,k'} 
\label{F1def}
\end{eqnarray}
and 
\begin{eqnarray} 
{\mathcal F}_2(k) \equiv {\mathcal F}_2(k;k',\omega) = 
{\mathcal F}_1(-k;k',\omega)
\label{F2def}
\end{eqnarray}
can be derived 
\begin{eqnarray} 
{\mathcal F}_1'(k) & = &\tilde{U}^2
\frac{\Theta(k'-k)}{\tilde{\omega}+k}- \tilde{U} \frac{{\mathcal
    F}_2(k)}{\tilde{\omega}+k } ,
\label{DGLF1} 
\\
{\mathcal F}_2'(k) & = & -\tilde{U}^2
\frac{\Theta(k'+k)}{\tilde{\omega}-k} + \tilde{U} \frac{{\mathcal
    F}_1(k)}{\tilde{\omega}-k } .
\label{DGLF2}
\end{eqnarray}
The boundary conditions are  ${\mathcal F}_1(-k_F)={\mathcal
  F}_2(k_F)= 0 $. 

We did not succeed in analytically solving this
system of differential equations for arbitrary $k$, $k'$, and small 
$|\omega|$. To determine the leading small $|\omega|$ behavior of
$g(\omega)$ and $h_1(\omega)$ we fortunately  only need ${\mathcal
  F}_1(k;k',\omega)$ for certain special combinations of the
arguments. 
From Eqs.\ (\ref{F1def}) and (\ref{gdef}) it follows that 
$g(\omega)$ is given by 
\begin{eqnarray}
g(\omega) = \left[
{\mathcal F}_1(\tilde{\omega};\tilde{\omega},\omega) 
+ \tilde{\mathcal V}_{\tilde{\omega},\tilde{\omega}} 
\right]^2 .
\label{gexpress}
\end{eqnarray}
The differential equation (\ref{DGLF2}) can be used to express
$h_1(\omega)$ Eq.\ (\ref{hdef}) in terms of ${\mathcal
  F}_1(k_F;\tilde{\omega},\omega)$\cite{wavefunctions}
\begin{eqnarray}
h_1(\omega) & = &
\left[ 1+ {\mathcal P} \! \! \! \! \! \! 
\int_{-k_F}^{k_F} d \, k 
\frac{{\mathcal F}_1(k;\tilde{\omega},\omega) 
+ \tilde{\mathcal V}_{k,\tilde{\omega}}}{\tilde{\omega} - k} 
\right]^2 \nonumber \\
& = &  \left[ 1+ \frac{1}{\tilde{U}} {\mathcal P} \! \! \! \! \! \! 
\int_{-k_F}^{k_F} d k {\mathcal F}_2'(k)
\right]^2 \nonumber \\
& = & \left[  1- \frac{1}{\tilde{U}}  
{\mathcal  F}_1(k_F;\tilde{\omega},\omega)  \right]^2 .
\label{hexpress}
\end{eqnarray}
We thus only have to evaluate 
${\mathcal F}_1(k_F;\tilde{\omega},\omega)$ and 
${\mathcal F}_1(\tilde{\omega};\tilde{\omega},\omega)$ in the limit 
$|\omega| \ll v_F k_F$. 

In a first step Eqs.\ (\ref{DGLF1}) and (\ref{DGLF2}) are solved for
$\omega = 0$. In this case a new variable $t(k) = \ln{|k|}$ can be
introduced and the equations can be decoupled by introducing
${\mathcal F}_{\pm}= {\mathcal F}_1 \pm {\mathcal F}_2$. The rest of
the calculation is straightforward and leads for $|k| \geq |k'|$ to
\begin{eqnarray}
{\mathcal F}_1(k;k',\omega=0) & = &\tilde{U} \Theta(k+k') \nonumber \\
&& \hspace{-1.5cm} -\frac{\tilde{U}}{2} 
  \left[ \left|\frac{k}{k_F} \right|^{- \tilde{U}} + 
           \, \mbox{sign}(k) \left|\frac{k}{k_F} \right|^{ \tilde{U}}
           \right] \left|\frac{k'}{k_F} \right|^{- \tilde{U}} . 
\label{F0calc}
\end{eqnarray}
For $|k'| > |k|$, $k$ and $k'$ have to be interchanged. 
If we set $k=k_F$, $k'=\tilde{\omega}$, and assume $|\omega| \ll v_F
k_F$, Eq.\ (\ref{F0calc})
simplifies to 
\begin{eqnarray}
{\mathcal F}_1(k_F;\tilde{\omega},0) & = &
\tilde{U} - \tilde{U}  \left|\frac{\omega}{v_F k_F}\right|^{-
  \tilde{U}} \nonumber \\
&& \hspace{-1.cm} = \tilde{U}^2 \ln{ \left|\frac{\omega}{v_F k_F}\right|} -
\frac{\tilde{U}^3}{2} \ln^2{ \left|\frac{\omega}{v_F k_F}\right|} 
+ \ldots \,\, . 
\label{F0spez}
\end{eqnarray}
The first few terms in the expansion in powers of $\tilde{U}$ can
easily be obtained by iterating the Lippmann-Schwinger equation
(\ref{LippSchwingK}) for the $K$ matrix. Unfortunately the quantity
presented in Eq.\ (\ref{F0spez}) is not quite what is needed to
calculate $h_1(\omega)$. An exact solution is difficult, but 
it is
straightforward to calculate the first terms in the $\tilde{U}$
expansion for
$\left[\tilde{K}(\omega)\right]_{k_F,\tilde{\omega}}$.    
For $|\omega| \ll v_F k_F$ this yields
\begin{eqnarray*}
{\mathcal F}_1(k_F;\tilde{\omega},\omega) 
 =  \tilde{U}^2 \ln{ \left|\frac{2 \omega}{v_F k_F}\right|} -
\frac{\tilde{U}^3}{2} \ln^2{ \left|\frac{2 \omega}{v_F k_F}\right|} 
+ \ldots \,\, .
\end{eqnarray*}
If we now assume that this expansion can be resummed as in Eq.\
(\ref{F0spez}) we obtain 
\begin{eqnarray}
h_1(\omega) = 
\left| \frac{2 \omega}{v_F k_F} \right|^{- 2 \tilde{U}} .
\label{hfinal}
\end{eqnarray}
For $k=\tilde{\omega}$ and $k'=\tilde{\omega}$ Eq.\ (\ref{F0calc}) gives 
\begin{eqnarray}
{\mathcal F}_1(\tilde{\omega};\tilde{\omega},0) & = &
\tilde{U} \Theta(\omega) - \frac{\tilde{U}}{2} \left[ 
\left|\frac{\omega}{v_F k_F}\right|^{- 2 \tilde{U}} +
\mbox{sign}(\omega) \right]  \nonumber \\
&& \hspace{-1.cm}  =  \tilde{U}^2 \ln{ \left|\frac{\omega}{v_F k_F}\right|} -
\tilde{U}^3 \ln^2{ \left|\frac{\omega}{v_F k_F}\right|} 
+ \ldots \,\, . 
\label{F0spez2} 
\end{eqnarray}
Iteration of Eq.\ (\ref{LippSchwingK}) for
$\left[\tilde{K}(\omega)\right]_{\tilde{\omega},\tilde{\omega}}$ leads
to 
\begin{eqnarray*}
{\mathcal F}_1(\tilde{\omega};\tilde{\omega},\omega)  = 
 \tilde{U}^2 \ln{ \left|\frac{ 2 \omega}{v_F k_F}\right|} -
\tilde{U}^3 \ln^2{ \left|\frac{2 \omega}{v_F k_F}\right|} 
+ \ldots \,\, .
\end{eqnarray*}
If we again assume that the logarithm can be summed to a power-law as
in Eq.\ (\ref{F0spez2}) we obtain the result for $g(\omega)$ presented
in Eq.\ (\ref{geval}).



\begin{thebibliography}{*}
\bibitem{spincite} Here we only consider the case of an equal number
  of electrons with up and down spin.
\bibitem{spinindependent} As the Green's and spectral function are
  independent of the spin direction $s$ we have suppressed the spin index.
\bibitem{disp} Here we neglect all the problems which might occur due
  to local minima and maxima in $\xi(k)$ or bandedges.
\bibitem{HnotHF} For the Hubbard model the Fock term vanishes. Thus
  we write $\rho^{H}$ instead of $\rho^{\rm HF}$.
\bibitem{missprint} The leading behavior of $K_{\rho}$ is given by
  $K_{\rho} = 1- U/(2 \pi v_F) $. Note the misprint in Eq.\
  (8) of Ref.\ [7].
\bibitem{wavefunctions} Note that
  $\tilde{\varphi}_{k}(j)/\tilde{\varphi}_{\tilde{\omega}}(j)$ 
  in Eq.\ (\ref{hdef}) has been replaced by 1.
\end{thebibliography}
\end{document}